\documentclass[reprint,amsmath,amssymb,prb,aps]{revtex4-2}
\usepackage{graphicx}
\usepackage{dcolumn}
\usepackage{bm}
\usepackage{orcidlink}
\usepackage[utf8]{inputenc}
\usepackage[mathlines]{lineno}
\usepackage{xcolor}
\usepackage{algpseudocode}
\usepackage{multirow}
\usepackage[caption=false]{subfig}
\begin{document}

\title{Bayesian Prior Construction for Uncertainty Quantification in First-Principles Statistical Mechanics}

\author{Derick E. Ober\orcidlink{0000-0001-7071-9406}}
\author{Sesha Sai Behara\orcidlink{0000-0003-2144-4586}}
\author{Anton Van der Ven}
\email{avdv@ucsb.edu}
\affiliation{Materials Department, University of California Santa Barbara}
\date{\today}

\begin{abstract}
First-principles statistical mechanics enables the prediction of thermodynamic and kinetic properties of materials, but is computationally expensive.
Many approaches require surrogate models to calculate energies within Monte Carlo or molecular dynamics simulations. 
Inexpensive surrogates such as cluster expansions enable otherwise intractable calculations by interpolating data from higher accuracy methods, such as Density Functional Theory (DFT). 
Surrogate models introduce uncertainty into downstream calculations, in addition to any uncertainty inherent to DFT calculations. 
Bayesian frameworks address this by quantifying uncertainty and incorporating expert knowledge through priors. 
However, constructing effective priors remains challenging. 
This work introduces and describes practical strategies for building Bayesian cluster expansions, focusing on basis truncation, hyperparameter selection, and ground state replication. 
We analyze multiple basis truncation schemes, compare cross-validation to the evidence-approximation for hyperparameter optimization, and provide methods to find and enforce ground-state-preserving models through priors. 
Additionally, we compare the uncertainties between different approximations to DFT (LDA, PBE, SCAN) against the uncertainty introduced with the use of cluster expansion surrogate models. 
These approaches are demonstrated on the BCC Li$_x$Mg$_{1-x}$ and Li$_x$Al$_{1-x}$ alloys, which are both of interest for solid-state Li batteries. 
Our results provide guidelines for constructing and utilizing Bayesian cluster expansions, thereby improving the transparency of materials modeling.
The approaches and insights developed in this work can be transferred to a wide range of cluster expansion surrogate models, including the atomic cluster expansion and related machine-learned interatomic potential architectures. 
\end{abstract}

\maketitle
\section{Introduction}

Statistical mechanics connects a material's thermodynamic and kinetic properties to its electronic structure.\cite{jaynes1957information,hill,allnatt2003atomic,van2018first}
Most statistical mechanics approaches require the evaluation of the energy of a large number of sampled microstates corresponding to excitations of atomic and electronic degrees of freedom. 
Ideally, the evaluation of the energy of each sampled microstate is done with a high-accuracy electronic structure method, such as Density Functional Theory (DFT). 
In practice, the number and scale of sampled microstates are too large to make high-accuracy electronic structure methods tractable as a primary tool to calculate energies. 
Instead, a computationally inexpensive surrogate model is used to evaluate microstate energies. 
Surrogate models predict energies for new microstates by interpolating and generalizing data from higher-accuracy methods. 

The cluster expansion is a mathematically rigorous surrogate model, ideally suited for statistical mechanics applications. 
It was first introduced by Sanchez et al \cite{sanchez1984generalized,laks1992efficient,de1994cluster} to address configurational degrees of freedom in alloys and was subsequently generalized for other degrees of freedom.\cite{drautz2004spin,thomas2013finite,thomas2018hamiltonians,natarajan2018machine,drautz2019atomic,van2018first,thomas2019machine,puchala2023casm,behara2024chemomechanics}
The approach prescribes how a complete and orthonormal set of basis functions of local degrees of freedom should be formulated for linear expansions of properties that emerge from a collection of interacting atoms.
The simpler cluster expansions, including the original alloy cluster expansion \cite{sanchez1984generalized} and the magnetic cluster expansions \cite{drautz2004spin}, are restricted to local perturbations relative to a high symmetry reference crystal structure. 
Recent extensions enable a description of atomic interactions in free space \cite{drautz2019atomic} and serve as a foundation with which to understand and interpret most machine-learned interatomic potential architectures. \cite{schutt2017schnet,unke2019physnet,gasteiger2020directional,anderson2019cormorant,thomas2018tensor,weiler20183d,satorras2021n,schutt2021equivariant,haghighatlari2022newtonnet,klicpera2021gemnet,tholke2022torchmd,brandstetter2021geometric,batzner20223,duval2023hitchhiker,batatia2025design}

Given infinite terms, a cluster expansion can exactly replicate the predictions of a higher-accuracy method. 
In practice, only a small number of terms are necessary to replicate a higher-accuracy method with minimal error. 
However, even high-accuracy methods such as DFT involve approximations and numerical tolerances. 
When a high-accuracy model is approximated with a surrogate model, existing uncertainty is compounded with additional uncertainty in the surrogate model. 
Uncertainty within the surrogate model will produce uncertainty in downstream predictions of thermodynamic properties \cite{natarajan2017symmetry,van2018first,cao2018use,van2020rechargeable,kadkhodaei2021cluster,behara2024chemomechanics,jung2025phase}. 
These uncertainties can be formally quantified and tracked within a Bayesian framework.\cite{Bishop2006}

Bayesian approaches to parameterizing linear surrogate models using high-accuracy training data exploit prior knowledge. 
This offers flexibility and a transparent method of regularizing surrogate models, but requires hyperparameters that need to be specified. 
Furthermore, while cluster expansions are formally exact if all basis functions are included, they necessarily need to be truncated. 
The specification of hyperparameters and the level of truncation require guiding principles. 
Finally, the ground states of a system are of central importance in determining the qualitative behavior of the system at finite temperature.  
A ``good" surrogate model that generalizes a high-accuracy method, such as DFT, should replicate the ground states while maintaining a low predictive error. Standard Bayesian methods do not recognize this qualitative ground state criterion. However, as recently shown,\cite{ober2024thermodynamically} ground state knowledge can be included as Bayesian priors to guide the construction of cluster expansion models and to perform uncertainty quantification.

Here, we analyze Bayesian methods in the context of constructing probability distributions for cluster expansion surrogate models. 
We use the Li$_x$Mg$_{1-x}$ and Li$_x$Al$_{1-x}$ alloy systems as models to systematically compare different approaches to determining Bayesian hyperparameters, truncation schemes, and ground state enforcement. 
While cross-validation is commonly used to guide truncation and hyperparameter selection,\cite{van2002automating} Bayesian approaches such as the evidence approximation,\cite{mackay1992bayesian,Bishop2006} and more specifically the relevance vector machine (RVM),\cite{tipping2001sparse} are found to be especially effective in specifying hyperparameters and model sparsification. 
This reinforces earlier work by Aldegunde et al \cite{ALDEGUNDE2016} who were the first to apply the RVM in the context of cluster expansion model construction. 

For consistent uncertainty quantification, all cluster expansion models drawn from a Bayesian posterior distribution should predict the same ground states.\cite{ober2024thermodynamically} 
The space spanned by the expansion coefficients of a cluster expansion naturally divides into cones of models that predict the same ground state sets. 
This property can be exploited in the formulation of Bayesian prior distributions that restrict the posterior distribution to cluster expansion models that predict the same ground state sets. 
Applied to the Li$_x$Al$_{1-x}$ alloy system, with its rich and diverse set of ordered ground state configurations, we demonstrate that standard Bayesian posterior distributions, such as the RVM, straddle a large number of ground state cones, thereby generating cluster expansion models that predict a wide variety of ground state sets. 
The use of prior distributions restricted to a particular ground state cone enables the construction of a posterior Bayesian distribution of models that predict the same ground state set. 
In this context, we introduce an approach to identify cluster expansion models that predict a prescribed ground state set. 
The approaches described and analyzed in this work can be transferred to other surrogate cluster expansion models, including the atomic cluster expansion, which is the foundation of machine learned interatomic potentials.\cite{drautz2019atomic}

\section{Methods}
\label{sec:methods}

\subsection{Cluster Expansion Surrogate Model}
An alloy cluster expansion \cite{sanchez1984generalized,de1994cluster} is a mathematically rigorous surrogate model that describes the dependence of the energy of a multi-component crystal as a function of the arrangement of its chemical species. 
As examples, we will focus on the Li-Mg and Li-Al binary alloys, which form BCC solid solutions and BCC intermetallic compounds. 
In specifying a decoration of $A$ and $B$ atoms on the sites of a parent crystal, such as BCC, each site $i$ of the crystal is assigned an occupation variable $s_i$ that is $+1$ if occupied by $A$ and $-1$ if occupied by $B$. 
A particular arrangement of $A$ and $B$ atoms on the BCC parent crystal structure can then be tracked by an array of occupation variables $\vec{s}=(s_1,\dots,s_i,\dots,s_N)$. 

A cluster expansion of the configuration dependence of the energy of a binary alloy is a sum of polynomial basis functions of occupation variables multiplied by weights according to
\begin{equation}
    E(\vec{s})=\sum_{c}V_{c}\Phi_{c}(\vec{s}).
    \label{eqn:cluster_expansion}
\end{equation}
The sum extends over all clusters of sites on the parent crystal, with the polynomial basis function associated with each cluster having the form
\begin{equation}
    \Phi_{c}(\vec{s})=\prod_{i\in c}s_{i}
\end{equation}
Each basis function is a product of occupation variables belonging to sites of a cluster $c$. 
The weights, $V_{c}$, are referred to as effective cluster interactions (ECI). 
Their numerical values depend on the parent crystal structure and the chemical nature of the alloy. 
As shown by Sanchez et al \cite{sanchez1984generalized}, the polynomial cluster basis functions $\Phi_{c}(\vec{s})$ form a complete and orthonormal basis in the space of alloy configurations for a particular definition of a scalar product. 
For a binary alloy having $N$ sites, there are 2$^N$ distinct arrangements of $A$ and $B$ atoms. 
There are also exactly 2$^N$ distinct clusters of sites, including the empty cluster along with $N$ point clusters, $N(N-1)/2$ pair clusters, etc.

\begin{figure}
    \centering
    \includegraphics[width=1.0\linewidth]{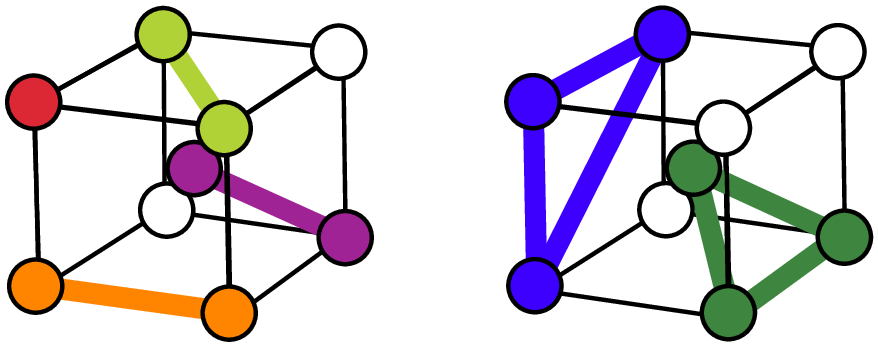}
    \caption{Bcc parent crystal structure with examples of a point cluster, pair clusters, and triplet clusters.}
    \label{fig:BCC_clusters}
\end{figure}

The underlying symmetry of the parent crystal structure reduces the number of independent ECI. 
For example, all nearest neighbor pair clusters of a parent BCC lattice will have the same ECI. 
All cluster functions $\Phi_{c'}(\vec{s})$ that can be mapped onto a prototype cluster function $\Phi_{c}(\vec{s})$ with a space group operation of the parent crystal can be grouped into an orbit of cluster functions $\Omega_{c}$. 
All cluster basis functions belonging to a common orbit have the same ECI in Eq.\ref{eqn:cluster_expansion}.
The cluster expansion (Eq. \ref{eqn:cluster_expansion}) can then be refactored according to
\begin{equation}
    E\left(\vec{s}\right)=\sum_{\Omega_{c}}V_{c}\left(\sum_{c'\in \Omega_{c}}\phi_{c'}\left(\vec{s}\right)\right)
    \label{eq:cluster_expansion2}
\end{equation}
The outer sum extends over each orbit of cluster functions, $\Omega_{c}$, while the inner sum extends over all cluster basis functions belonging to that orbit. 
Upon normalizing this expression by the number of primitive unit cells of the parent crystal structure, $N_{u}$, the energy per unit cell becomes
\begin{equation}
    e\left(\vec{s}\right)=\sum_{\Omega_{c}}w_c\xi_{c}\left(\vec{s}\right)
    \label{eqn:cluster_expansion_normalized}
\end{equation}
where 
\begin{equation}
    \xi_{c}\left(\vec{s}\right)=\frac{1}{N_{u}m_{c}}\left(\sum_{c'\in \Omega_{c}}\phi_{c'}\left(\vec{s}\right)\right)
    \label{eqn:correlation_function}
\end{equation}
are referred to as correlation functions,\cite{sanchez1978fcc} and where
\begin{equation}
    w_c=m_cV_c.
\end{equation}
In these expressions, $m_{c}$ is the multiplicity of symmetrically equivalent cluster basis functions per primitive unit cell. 
The collection of all the distinct correlation functions can be represented as a vector, i.e. $\vec{\xi}^{\mathsf{T}}_{\vec{s}}=(\xi_{0},\xi_{1},\dots,\xi_{c},\dots)$. 
This vector acts as a ``fingerprint" describing a given configuration $\vec{s}$ of $A$ and $B$ atoms over the sites of the parent crystal. 
The conjugate set of ECI (multiplied by their multiplicities) can similarly be represented as a vector $\vec{w}^{\mathsf{T}}=(w_0,w_1,\dots,w_c,\dots)$.
Eq. \ref{eqn:cluster_expansion_normalized} can then be recast as a scalar product between the correlation vector, $\vec{\xi}_{\vec{s}}$, and the ECI vector, $\vec{w}$, as
\begin{equation}
e(\vec{s})=\vec{\xi}^{\mathsf{T}}_{\vec{s}}\vec{w}
\label{eqn:cluster_expansion_as_scalar_product}
\end{equation}
A cluster expansion is an exact description of the configuration-dependent energy of a multi-component crystal since it is formulated in terms of a complete and orthonormal basis in configuration space.\cite{sanchez1984generalized,de1994cluster}
Hence, each alloy system on a particular parent crystal structure 
has a unique ECI vector $\vec{w}$ that enables the calculation of the energy of any configuration $\vec{s}$ according to Eq. \ref{eqn:cluster_expansion_as_scalar_product}.


\subsection{Training Data}


A variety of quantum mechanical methods can be used to calculate the energy of a chemical ordering on a parent crystal structure from first principles. Most approaches derive from density functional theory (DFT). DFT recasts the energy as determined by the many-body Schrödinger equation into a functional of the electronic charge density and an external potential. 
It is formally exact, but approximations are necessary to make it a practical approach. 
Commonly used approximations include the local density approximation (LDA), the generalized gradient approximation (GGA) (which includes parameterizations such as PW91, PBE ), and meta-GGA, including SCAN. \cite{kohn1965self,perdew1991electronic,perdew1996generalized,sun2015strongly}
Each approximation to DFT predicts a self-consistent set of energies for every chemical ordering on a parent crystal structure of a particular alloy. 
Therefore, there is a unique ECI vector $\vec{w}$ that defines a cluster expansion of the energy of configurations according to Eq. \ref{eqn:cluster_expansion_as_scalar_product} for each DFT approximation. 

Within a particular approximation to DFT (e.g. LDA, GGA, etc.), errors are incurred when calculating the energy of a crystal using numerical methods.\cite{lejaeghere2016reproducibility} 
Numerical errors arise from the necessary truncation of the electron orbital basis and the use of a k-point grid in reciprocal space to integrate the contribution from the electronic band structure to the total energy.
Hence, while the true energy of a configuration $\vec{s}$ can be calculated according to Eq. \ref{eqn:cluster_expansion_as_scalar_product}, using an ECI vector $\vec{w}$ for a particular approximation to DFT, a numerical calculation of the energy of a configuration, $t_{\vec{s}}$, will differ from Eq. \ref{eqn:cluster_expansion_as_scalar_product}.
Instead of an equality, the $t_{\vec{s}}$ as calculated with a numerical approach will be tainted with numerical errors, $\epsilon_{\vec{s}}$, such that 
\begin{equation}
    t_{\vec{s}}=\vec{\xi}_{\vec{s}}^{\mathsf{T}}\vec{w}+\epsilon_{\vec{s}}
    \label{eqn:cluster_expansion_plus_noise}
\end{equation}
A common assumption is that the noise $\epsilon_{\vec{s}}$ on $t_{\vec{s}}$ due to numerical errors is identically and independently distributed (iid). 
The noise of each configuration $\vec{s}$ is then drawn from the same distribution and is not correlated between configurations. 
The noise distribution on numerically calculated energies can further be assumed to be a zero-centered Gaussian, characterized by a precision $\beta$: i.e. $ p(\epsilon_{\vec{s}} | 0,\beta^{-1}) = \mathcal{N}(\epsilon_{\vec{s}} | 0,\beta^{-1})$ where $\mathcal{N}$ represents a normal distribution with mean $\mu=0$ and variance $\sigma_t^{2}=\beta^{-1}$. 
With these assumptions about the numerical noise of a first-principles energy calculation, the probability of numerically calculating an energy $t_{\vec{s}}$ given the true ECI vector, $\vec{w}$, for a particular alloy and approximation to DFT is then
\begin{equation}
    \mathcal{N} ( t_{\vec{s}}|\vec{\xi}_{\vec{s}}^{\mathsf{T}}\vec{w},\beta^{-1} ) = \left( \frac{\beta}{2\pi} \right)^{1/2}\text{exp} \left\{ -\frac{\beta}{2} (t_{\vec{s}} - \vec{\xi}_{\vec{s}}^{\mathsf{T}} \vec{w})^2 \right\}
\end{equation}

First-principles calculations of the energy of a crystal in a particular configuration are computationally expensive and are restricted to relatively small periodic orderings. Therefore, it is only possible to calculate the energies of a very small fraction of the $2^N$ possible configurations of a binary alloy on a parent crystal with $N$ sites (with $N \rightarrow \infty$ in the thermodynamic limit). 
The formation energies of $n \ll 2^{N}$ configurations (on the order of several hundred to a thousand) can then be collected in a vector $\vec{t}$. Eq. \ref{eqn:cluster_expansion_plus_noise} then generalizes to
\begin{equation}
\vec{t} = X\vec{w} + \vec{\epsilon}.
\end{equation} 
The matrix $X$ is composed of $n$ correlation row vectors $\vec{\xi}^{\mathsf{T}}_{\vec{s}}$, while $\vec{\epsilon}$ is a vector of random variables that each are drawn from a Gaussian distribution with mean $\mu=0$ and variance $\sigma_t^2=\beta^{-1}$.
When the iid noise $\epsilon$ is taken from a normal (Gaussian) distribution, the likelihood function is a product of univariate distributions:
\begin{equation}
P(\vec{t}|X, \vec{w}, \beta) = \prod^{n}_{i} \mathcal{N} (t_i|X_i\vec{w},\beta^{-1})
\label{eqn:likelihood_function}
\end{equation}
where $X_i$ represents a row of the correlation matrix $X$.
Equation \ref{eqn:likelihood_function} measures the likelihood that the numerically calculated first principles energies for $n$ configurations on a parent crystal structure would have the values collected in the vector $\vec{t}$, given that $\vec{w}$ is the true alloy ECI vector and that the numerical method used to calculate $\vec{t}$ has an error drawn from a Gaussian centered at zero and having a variance of $\sigma_t^2=\beta^{-1}$.

\subsection{Bayesian inference of expansion coefficients}
\label{sec:bayesian_inference_of_expansion_coefficients}
While an alloy on a particular parent crystal has a unique ECI vector for each flavor of DFT, this vector is not known. 
Eq. \ref{eqn:likelihood_function} assumes knowledge of the complete ECI vector $\vec{w}$ and measures the probability that a particular first-principles numerical method, with a precision $\beta$, will generate the set of energies $\vec{t}$ for $n$ configurations. 
Since $\vec{w}$ is not known for a real alloy, it is desirable to invert the problem and estimate $\vec{w}$ given the $n$ training data collected in the vector $\vec{t}$.
To this end, it is necessary to construct a probability distribution for the ECI vector $\vec{w}$ given the training data $\vec{t}$.
Such a probability distribution can be formulated in terms of the likelihood distribution, Eq. \ref{eqn:likelihood_function}, using Bayes' theorem.\cite{Bishop2006,Mueller2009,kristensen2014bayesian,ALDEGUNDE2016} 
In its most general form, Bayes' theorem relates the conditional probability of a random variable ``$A$" given ``$B$" to the conditional probability of ``$B$" given ``$A$" according to
\begin{equation}
P(A|B) = \frac{P(B|A)P(A)}{P(B)}.  
\label{eqn:bayes_theorem}
\end{equation}
Applied to the probability of a cluster expansion, Bayes' theorem becomes \cite{Mueller2009}
\begin{equation}
    P(\vec{w}|\vec{t},X,\beta) = \frac{P(\vec{t}|X,\vec{w},\beta)P(\vec{w})}{P(\vec{t}|X,\beta)}.
\label{eqn:bayes_theorem_contextualized}
\end{equation}
In this relation, $P(\vec{t}|X,\vec{w},\beta)$ is the likelihood function introduced in Eq. \ref{eqn:likelihood_function}. 
The $P(\vec{w})$ is a prior distribution on the ECI vector $\vec{w}$. Within a Bayesian framework, the prior can encode expert knowledge about the ECI before data is observed. 
The denominator, $P(\vec{t}|X,\beta)$, serves as a normalization factor and does not need to be addressed explicitly. 
The posterior distribution for $\vec{w}$ makes it possible to sample cluster expansion models that are consistent with the training data $\vec{t}$.\cite{kristensen2014bayesian,ALDEGUNDE2016,ober2024thermodynamically} 
So far, $\vec{w}$ represents the full ECI vector, which in the thermodynamic limit of a crystal (i.e. $N\rightarrow \infty$) becomes infinitely large.

\subsection{Choosing a prior distribution}
\label{sec:prior_distributions}
Several categories of prior knowledge can be encoded in $P(\vec{w})$ of Eq. \ref{eqn:bayes_theorem_contextualized}.
These can be used to formally justify truncation and regularization and to constrain the qualitative predictions of cluster expansions sampled from the posterior distribution. 
A common choice for the prior probability distribution of an individual expansion coefficient, $w_c$, centered around a mean of $\left<w_c\right>_p=0$, has the form \cite{Bishop2006}
\begin{equation}
    P(w_c|\alpha_c) \sim \text{exp} \{-\frac{\alpha_c}{2}|w_c|^p\},
    \label{eqn:prior1}
\end{equation}
which has a normalization factor of
\begin{equation}
    \frac{p}{2\Gamma(1/p)}\left(\frac{\alpha_{c}}{2}\right)^{1/p}
\end{equation}
 In this expression, $\Gamma$ is the gamma function, while $p$ determines the nature of the distribution. 
Setting $p=2$ yields a normal (Gaussian) distribution, while $p=1$ generates the Laplacian distribution. 
Other values of $p$, (e.g. $p=1/2$) are also possible. 
The $\alpha_c$ is the precision of the prior distribution and for $p=2$ is equal to the inverse of the variance, i.e. $\alpha_c=1/\sigma_c^{2}$. 
Assuming that the prior distributions on each ECI $w_c$ are independent of each other, the prior distribution on the ECI vector, $\vec{w}$, can be generated as the product over individual distributions
\begin{equation}
    P(\vec{w}|\vec{\alpha})=\prod_c P(w_c|\alpha_c)
\end{equation}
where $\vec{\alpha}$ is a vector that collects the precisions, $\alpha_c$, for each ECI, $w_c$.
There are scenarios where correlations between different ECIs are meaningful \cite{Mueller2009}.

\begin{figure}
    \centering
\includegraphics[width=1\linewidth,height=1\textheight,keepaspectratio]{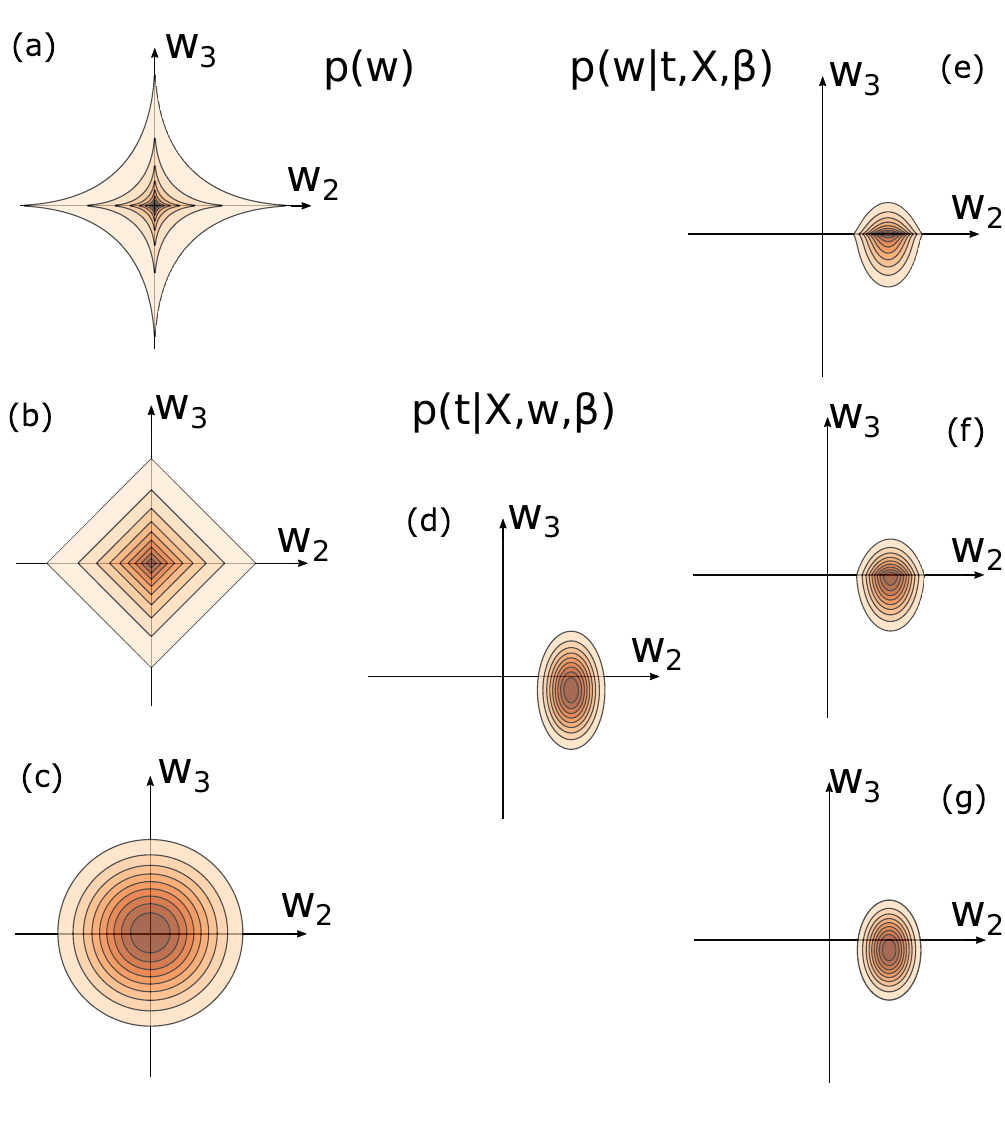}
    \caption{Contour plots of possible prior probability distributions (a-c). Each prior distribution (a-c) can be multiplied with the likelihood function (d) to produce a corresponding posterior distribution (e-g), respectively. The prior distributions have (a) $p=1/2$, (b) $p=1$ and (c) $p=2$. A value of $p=2$ produces a typical Gaussian contour, shown in (c). As $p \to 0$, the probability density is shifted closer towards the axes, away from the interior, off-axis space.}
    \label{fig:potential_contours}
\end{figure}

Interactions between the different chemical species of an alloy tend to be short-ranged in the absence of unscreened electrostatic forces and large inhomogeneous strain fields.\cite{laks1992efficient,wolverton2000first,wolverton2000short,wang2023generalization,vandewalle_CeO,behara2024chemomechanics} 
This motivates a truncation of $\vec{w}$, limiting the cluster expansion to small and short-ranged clusters. 
The truncation of the cluster expansion can be encoded in $P(\vec{w})$ by letting the precision $\alpha_c$ for the truncated clusters go to infinity ({\it i.e.} $\sigma_c \rightarrow 0$). 
The prior distribution for that expansion coefficient, $w_c$, then reduces to that of a $\delta$-function distribution, centered at zero.
The ECIs that are not truncated have finite precisions, $\alpha_c$, which serve to regularize the values of $w_c$ in the posterior distribution. 
The manner in which they constrain the size of $w_c$ depends on the value of $p$, as illustrated schematically in  Figure \ref{fig:potential_contours}(a), (b), (c) for a two-dimensional ECI space.

Other forms of prior knowledge can be incorporated to construct a posterior distribution on the ECI (Eq. \ref{eqn:bayes_theorem_contextualized}).
Qualitative predictions of a cluster expansion depend on the region of ECI space from which the ECI vector $\vec{w}$ is sampled. 
For example, each cluster expansion vector $\vec{w}$ predicts a particular set of ground state orderings, corresponding to the subset of configurations that have the lowest energy among all other configurations.\cite{inden,ober2024thermodynamically} 
Cluster expansions that all predict the same set of ground states reside in ray-bounded ``cones" in ECI space that radiate out from the origin.\cite{ober2024thermodynamically} 
\begin{figure}
    \centering
\includegraphics[width=0.7\linewidth,height=0.6\textheight,keepaspectratio]{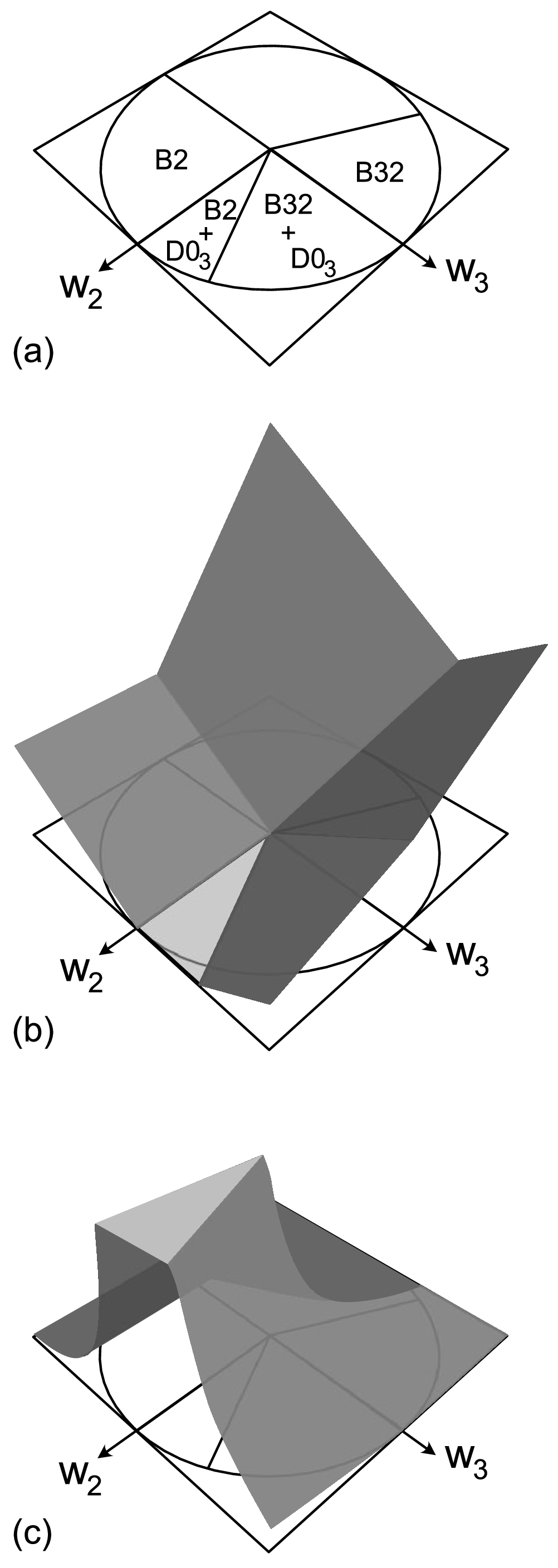}
    \caption{(a): Unique ground state sets for BCC, using only first and second nearest-neighbor pair interactions. Each cone corresponds to domains in ECI space in which a particular set of ground states are predicted. The ``empty" cone predicts a miscibility gap (phase separation). A cluster expansion with only first and second nearest-neighbor pair interactions predicts D0$_3$ as being stable at both $x=\frac{1}{4}$ and $x=\frac{3}{4}$. (b): The surface $\eta(\vec{w})$ of Eq. \ref{eqn:continuous_gsa_first} as a function of $w_2$ and $w_3$. (c): The negative exponential $\text{exp}\{-\eta(\vec{w})\}$. Subfigures b and c are enforcing the BCC ground state set of [B2,D0$_3$].}
    \label{fig:masking_function}
\end{figure}
As a simple example, consider a cluster expansion for the BCC parent crystal structure containing only the first and second nearest neighbor pair interactions, $w_2$ and $w_3$ (the constant and point interactions are labeled $w_0$ and $w_1$). 
Figure \ref{fig:masking_function}(a) shows different ground state cones in the space spanned by $w_2$ and $w_3$. 
Any ECI vector $\vec{w}=(w_2,w_3)^{\mathsf{T}}$ sampled from the same cone will predict the same set of ground states. 
For a first and second nearest neighbor cluster expansion model of the BCC parent crystal structure, the only possible ground states are B2, B32 and D0$_3$.\cite{inden} 
In general, the possible ground state sets and their associated domains depend on the specific cluster basis truncation. 
Severe truncation, leading to a small number of ECI, restricts the types of ordered phases that can be ground states. 
When a particular truncation of a cluster expansion fails to predict a the correct set of ground states, additional basis clusters may need to be added. 

\begin{figure}
    \centering
\includegraphics[width=0.7\linewidth,height=0.7\textheight,keepaspectratio]{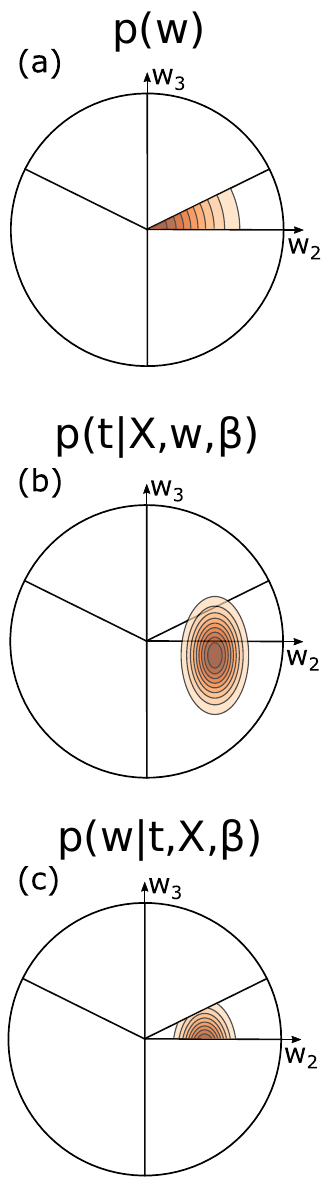}
    \caption{Multiplying the ground state constrained prior (a) with the likelihood function (b) produces a posterior distribution (c) that assigns higher probability to models that agree with observed data, but zero probability to models that do not match imposed ground states. 
    }
    \label{fig:BCC_l2_cone_prior}
\end{figure}

If there is a high degree of certainty about the ground states of a system, a prior distribution can be formulated that favors models residing within the correct ground state cone. 
To construct such a prior distribution, it is convenient to introduce a masking function, $\eta(\vec{w})$, defined as being zero within the correct ground state cone and monotonically increasing as $\vec{w}$ strays away from that domain.\cite{ober2024thermodynamically}
An example of a masking function is shown in Figure \ref{fig:masking_function}(b) for the first and second nearest neighbor cluster expansion of the BCC parent, where the cone corresponding to the [B2,D03] ground state set is filtered. 
A prior distribution of the form
\begin{equation}
    P(\vec{w}) \propto \text{exp} \{-\gamma \eta(\vec{w})\}
    \label{eqn:prior_general}
\end{equation}
will then bias the desired cone, with $\gamma$ serving as a ``tuning parameter" of the bias.
This prior distribution places higher probability density on models that accurately reproduce the desired ground states, with the parameter $\gamma$ controlling the intensity of this prior. 
If $\gamma \to \infty$, the prior reduces to a uniform probability density on models that correctly replicate the targeted ground state set, and zero probability density on models that predict a different set of ground states.
Combining Eq. \ref{eqn:prior_general} with a Gaussian mean-zero prior produces a prior distribution as illustrated in Figure \ref{fig:BCC_l2_cone_prior}(a), where probability density is zero outside of the target ground state domain. 
The product between this prior distribution and the likelihood function in Figure \ref{fig:BCC_l2_cone_prior}(b) produces a restricted posterior distribution as shown in Figure \ref{fig:BCC_l2_cone_prior}(c).


\subsection{The posterior distribution}

The various prior distributions, including those that encode truncation, regularization and bias for a particular ground state cone, can be combined with the likelihood function, Eq. \ref{eqn:likelihood_function}, to generate a posterior distribution of the form
\begin{equation}
    P(\vec{w}|\vec{t},X,\vec{\alpha},\beta,\gamma) \sim \text{exp} \{-\beta H(\vec{w},\vec{\lambda},\delta)\}
    \label{eqn:posterior_distribution}
\end{equation}
where $H(\vec{w},\vec{\lambda},\delta)$ is a cost function given by
\begin{equation}
    H(\vec{w},\vec{\lambda},\delta) = \frac{|\vec{t}-X\vec{w}|_2^2}{2} + \sum_{c}\frac{\lambda_c |w_c|^p}{2} + \delta \eta(\vec{w}).
    \label{eqn:generalized_kost_function}
\end{equation} 
In this expression, $\vec{\lambda}=\vec{\alpha}/\beta$ (i.e. $\lambda_c = \alpha_c/\beta$) and $\delta = \gamma/\beta$. 

The posterior distribution, Eq. \ref{eqn:posterior_distribution}, has the form of a Boltzmann factor from statistical mechanics with the components of the ECI vector, $\vec{w}$, serving as degrees of freedom and the cost function $H(\vec{w},\vec{\lambda},\delta)$ acting as a Hamiltonian. 
The normalization constant of the posterior distribution 
\begin{equation}
    Z(\beta,\vec{\lambda},\delta)=\int_{\vec{w}}\text{exp} \{-\beta H(\vec{w},\vec{\lambda},\delta)\}d\vec{w}
    \label{eqn:partition_functon}
\end{equation}
is similar to a partition function from statistical mechanics.
In the above expressions, $\beta$, which, in the absence of truncation errors, measures the precision on the training data, can then be viewed as an inverse temperature. 
In some cases, the posterior can be calculated analytically. 
For example, when $p=2$ and $\delta=0$, the cost function is parabolic in terms of $\vec{w}$ and the posterior distribution becomes Gaussian. 
The normalization factor, Eq. \ref{eqn:partition_functon}, of a multi-variable Gaussian posterior distribution can be integrated in closed form.\cite{Bishop2006} 
In general, though, this is not possible and Monte Carlo sampling of the ECI vector $\vec{w}$ from the posterior distribution is necessary.\cite{ober2024thermodynamically}
To ensure that the posterior can always be normalized, it may be necessary to use a regularizing prior distribution of the form of Eq. \ref{eqn:prior1}. 

As is commonly done in statistical mechanics,\cite{hill,puchala2025casmMC} a free energy potential can be defined in terms of the normalizing constant (i.e., the partition function) according to
\begin{equation}
    -\beta\Lambda=\ln Z(\beta,\vec{\lambda},\delta)
    \label{eqn:free_energy}
\end{equation}
The first derivatives of $\Lambda$ with respect to the hyperparameters, $\lambda_c=\alpha_c/\beta$ and $\delta=\gamma/\beta$, then yield expectation values of their conjugate variables. 
For example, 
the derivative of $\Lambda$ with respect to $\delta$ yields 
\begin{equation}
    \left(\frac{\partial \Lambda}{\partial \delta}\right)_{\beta,\lambda_c}=\left<\eta\right>, 
    \label{eqn:expectation_of_masking_function}
\end{equation}
the expectation value of the masking function.

\subsection{Ground state cones in ECI space}
\label{sec:finding_cone}
\begin{figure}
    \centering
    \includegraphics[width=1\linewidth]{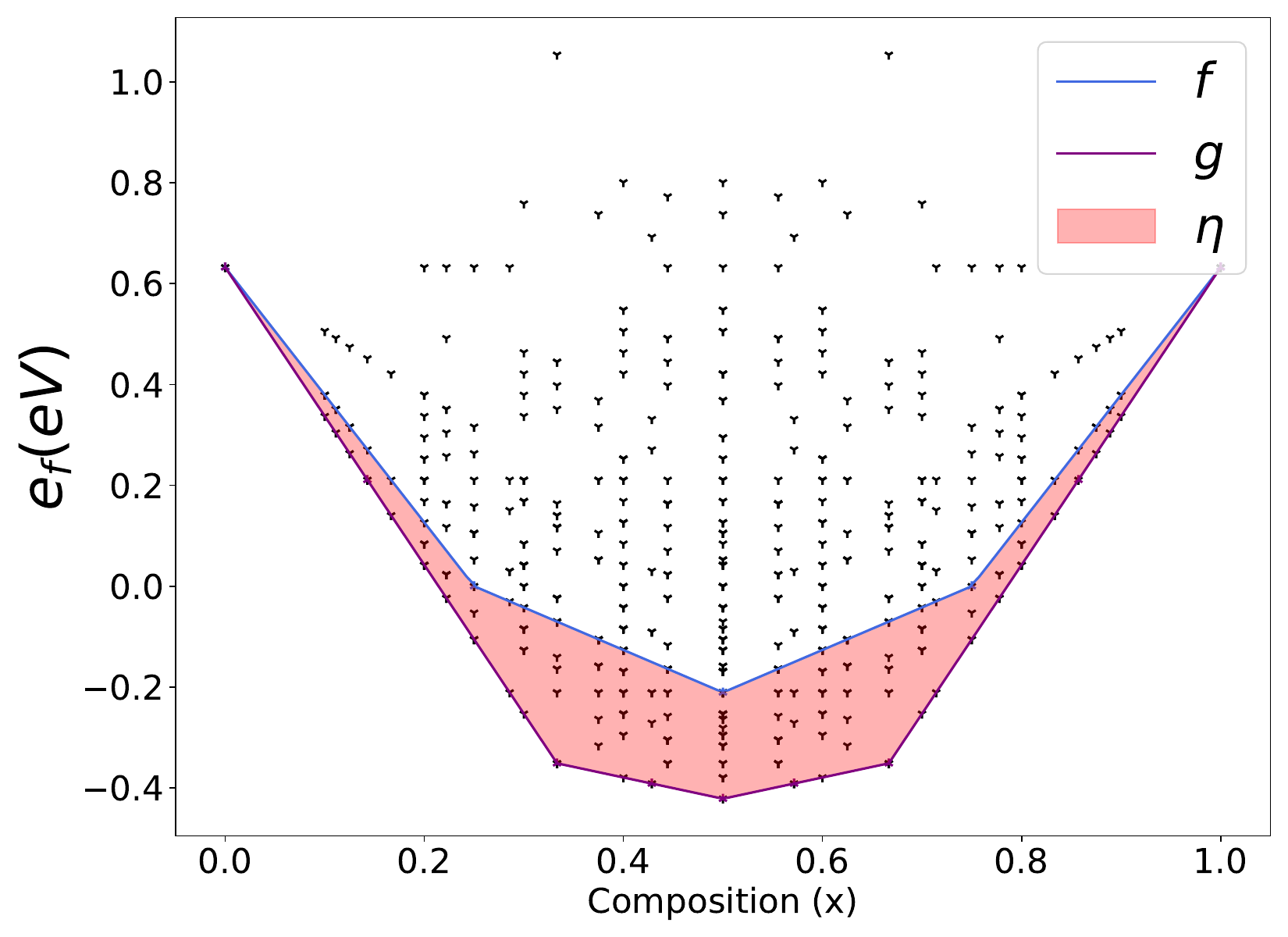}
    \caption{Visualization of a ground state masking function $\eta(\vec{w})$ from Eq. \ref{eqn:continuous_gsa_first}. The value of $\eta$ is zero when $f=g$, and positive otherwise. As $f$ and $g$ become more different, the value of $\eta$ increases.}
    \label{fig:envelope}
\end{figure}

The ground-state-informed prior of Eq. \ref{eqn:prior_general} can enforce ground state replication in sampled ECI vectors, but requires a definition for the masking function $\eta(\vec{w})$. 
To serve its role as introduced in Eq. \ref{eqn:prior_general}, the masking function, $\eta(\vec{w})$ should be zero for ECI vectors within the desired ground state domain, and increase as $\vec{w}$ moves away from that domain. 
The function should also be continuous across all of ECI space. 
One definition for $\eta(\vec{w})$ that satisfies these requirements is:\cite{ober2024thermodynamically} 
\begin{equation}
     \eta(\vec{w}) =  \int \left[f(\vec{w},x) - g(\vec{w},x)\right]dx
     \label{eqn:continuous_gsa_first}
\end{equation}
where $g(\vec{w},x)$ is the lower convex hull to the energies of all configurations as calculated with the ECI vector $\vec{w}$ and  $f(\vec{w},x)$ is the piece-wise linear function connecting the energies of the targeted ground states. 
Unlike $g(\vec{w},x)$, $f(\vec{w},x)$ does not have to be convex. 
A graphical interpretation of $\eta(\vec{w})$ is shown in Figure \ref{fig:envelope} as the red area between $f(\vec{w},x)$ and $g(\vec{w},x)$. 
To calculate $\eta(\vec{w})$, it is necessary to enumerate a large number of configurations, $\vec{s}$, using a combination of systematic approaches \cite{hart2008algorithm} and Monte Carlo annealing simulations to determine the ground states for the ECI vector $\vec{w}$.\cite{puchala2025casmMC} 

While the masking function $\eta(\vec{w})$ is designed to bias the prior probability distribution Eq. \ref{eqn:prior_general} to a specific ground state cone, it also offers a way of finding ECI vectors $\vec{w}$ that replicate the targeted ground states. 
This is important to determine whether a particular level of truncation of the cluster expansion is able to replicate the desired ground states and to identify a starting cluster expansion model for Monte Carlo sampling from a posterior distribution for $\vec{w}$. 
The determination of a point within a particular ground state cone can be achieved with gradient descent, provided that $\eta(\vec{w})$ is differentiable.
For any $\vec{w}$, the negative gradient of Eq. \ref{eqn:continuous_gsa_first} is 
\begin{equation}
    -\nabla_w \eta(\vec{w}) = \nabla_w  \int g(\vec{w},x)dx - \nabla_w  \int f(\vec{w},x)dx 
\end{equation}
which will point in the direction of the target cone in ECI space.
%
For a simple binary system, the integral along the composition axis $x$ of the function $f$, which is piece-wise linear, becomes 
\begin{equation}\label{binary_trapezoid}
    \int f(\vec{w},x)dx = \sum_{s \in S} \frac{1}{2}  (\vec{\xi}^{\mathsf{T}}_{s+1}\vec{w}+\vec{\xi}_s^{\mathsf{T}}\vec{w})(x_{s+1}-x_s)
\end{equation}
where $S$ represents the set of target ground states (ordered by increasing composition $x$ for a binary system). 
The vector $\vec{\xi}_{s}$ is the correlation vector of structure $s$, and $x_{s}$ is the composition of configuration $s$. 
The gradient of this integral with respect to $\vec{w}$ becomes 
\begin{equation}
    \nabla_w \int f(\vec{w},x) = \sum_{s \in S} \frac{1}{2} (\vec{\xi}^{\mathsf{T}}_{s+1}  + \vec{\xi}^{\mathsf{T}}_s ) (x_{s+1} - x_s) 
\end{equation}
Therefore, the direction that minimizes the metric $\eta(\vec{w})$ in a binary compound is 
 \begin{equation}
     \sum_{s \in S} (\vec{\xi}^{\mathsf{T}}_{s+1}  + \vec{\xi}^{\mathsf{T}}_s ) (x_{s+1} - x_s) - \sum_{s' \in S'}  (\vec{\xi'}^{\mathsf{T}}_{s'+1}  + \vec{\xi'}^{\mathsf{T}}_s ) (x'_{s+1} - x'_s) 
     \label{eqn:gradient}
 \end{equation}
where the primed (unprimed) quantities come from the integral of $f$ ($g$). 
The factor of 1/2 is dropped because it does not change the direction of gradient descent. 
This gradient can be evaluated at any point in the ECI space (except possibly for points exactly along a domain boundary). 
The gradient of $\eta({\vec{w}})$ with respect to $\vec{w}$ in higher-dimensional composition spaces is derived in Appendix \ref{app:gradient_masking_function}.

One approach to numerically identify an ECI vector $\vec{w}$ that resides within the desired ground state cone is to perform gradient descent to a starting vector using Eq. \ref{eqn:gradient}. 
This procedure can fail if the negative gradient points directly through the origin, resulting in the zero vector as a trivial solution. 
To minimize the risk of this occurring, gradient descent can be performed from multiple starting points, e.g. each axis in ECI space (the positive and negative unit vectors of each axis). 
If the truncated cluster expansion model is capable of replicating the targeted ground states, the resulting ECI vectors should line the perimeter of the ground state cone. 
The mean of these vectors will then yield a vector inside the ground state cone. 
When none of the gradient descent paths land in the targeted ground state cone, additional basis functions (corresponding to new ECI axes) may need to be added to give the model more flexibility in replicating the desired ground states.

\subsection{Hypothesis testing}
\label{sec:hypothesis_testing}
When performing uncertainty quantification on downstream predictions, it is often desirable that each sampled ECI vector predicts the same set of ground states.\cite{ober2024thermodynamically}
This can be achieved by letting $\gamma = \beta \delta \rightarrow \infty$ in the prior distribution of Eq. \ref{eqn:prior_general} where $\gamma$ is conjugate to a masking function $\eta(\vec{w})$ that is zero whenever $\vec{w}$ predicts the target ground states. 
In this limit, the prior distribution on $\vec{w}$ is non-zero in the cone of ECI space corresponding to the targeted ground states and zero everywhere else.

The ability to carve up ECI space into domains characterized by distinct qualitative behavior opens the door to hypothesis testing. 
The probability that a cluster expansion model resides in a constrained subdomain of ECI space relative to the full space, given a particular set of regularizing hyperparameters $\vec{\lambda} = \vec{\alpha}/\beta$ can be expressed as
\begin{equation}
    P(\mathcal{M})=\frac{\int_{\vec{w}}\text{exp}\{ -\beta H(\vec{w},\vec{\lambda},\delta=\infty)\} d\vec{w}}{\int_{\vec{w}}\text{exp}\{-\beta H(\vec{w},\vec{\lambda},\delta = 0)\}d\vec{w}}
    \label{eq:bayes_factor}
\end{equation}
where the integral in the numerator is restricted to the subdomain in which the masking function $\eta(\vec{w})$ is equal to zero, while the integral in the denominator extends over all of ECI space. 
These integrals are equal to partition functions, Eq. \ref{eqn:partition_functon}, for $\delta=\infty$ and $\delta=0$, respectively, and can be expressed in terms of a difference in free energy according to
\begin{equation}
    P(\mathcal{M})=\frac{Z(\beta,\vec{\lambda},\delta=\infty)}{Z(\beta,\vec{\lambda},\delta = 0)}=\text{exp}\{-\beta[\Lambda(\delta=\infty)-\Lambda(\delta=0)]\}
    \label{eq:model_probability}
\end{equation}
using Eq. \ref{eqn:free_energy}.
The difference in free energy can in turn be calculated by integrating Eq. \ref{eqn:expectation_of_masking_function} from $\delta=0$ to $\delta=\infty$
\begin{equation}
    \Lambda(\delta=\infty)-\Lambda(\delta=0) = \int_{0}^{\infty} \left<\eta\right>(\delta)d\delta
    \label{eq:free_energy_integral}
\end{equation}
The average $\left<\eta\right>(\delta)$ can be calculated with Monte Carlo simulations that sample $\vec{w}$ according to the posterior distribution, Eq. \ref{eqn:posterior_distribution}, and evaluate the masking function as $\delta$ is increased from 0 to $\infty$.
This enables a comparison of the probability of different ground state domains in ECI space by inserting Eq. \ref{eq:free_energy_integral} into Eq. \ref{eq:model_probability}.

\subsection{Choosing hyper parameters}

The Bayesian posterior distribution of the ECI vector $\vec{w}$ is a function of the hyperparameters $\vec{\alpha}$, $\beta$ and $\gamma$. 
There are two general frameworks to guide the selection of the $\vec{\alpha}$ and $\beta$ in the absence of any ground state bias (i.e. $\gamma=0$). 
One relies on the optimization of a cross-validation score, while another, referred to as the evidence approximation, selects hyper parameters that maximize $P(\vec{\alpha},\beta|\vec{t})$, the probability distribution of the hyper parameters given the training data $\vec{t}$. 
These are reviewed in the next two sections.

\subsubsection{Minimization of the cross-validation score}
\label{sec:hyperparameters_through_cv_minimization}

A subset of hyperparameters can be estimated by minimizing a $k$-fold cross-validation score. 
To this end, the DFT data is divided into $k$ equally sized folds with one fold serving as a test set and the remaining $k$-1 folds used in the likelihood function. 
The cross-validation (CV) error is defined as the root-mean-squared error between the predicted and DFT energies of the test set. 
An average cross-validation error is obtained by repeating this $k$ times, with each fold treated as a test set. 
The cluster expansion used to calculate the cross-validation score is the maximum $a$ $posteriori$ (MAP) estimate corresponding to the $\vec{w}$ that maximizes the posterior distribution, $P(\vec{w}|\vec{t},\vec{\alpha},\beta,\gamma)$, for fixed $\vec{\alpha}$ and $\beta$ (assuming $\gamma=0$).
Maximizing the posterior distribution, Eq. \ref{eqn:posterior_distribution}, is equivalent to minimizing the cost function $H(\vec{w},\vec{\lambda},\gamma=0)$ with respect to $\vec{w}$. 
Hence, for each $\vec{\lambda}=\vec{\alpha}/\beta$, there is a MAP vector $\vec{w}$ that minimizes the cost function and that can be used to calculate a CV-error. 
The sought-after hyperparameters are the $\vec{\lambda}$ that minimize the cross-validation score. 
While the CV-error can be minimized with respect to all elements of $\vec{\lambda}=(\dots,\lambda_c,\dots)$, this is generally not practical and the hyperparameters, $\lambda_c$, for non-truncated basis functions are simply set equal to a common value $\lambda = \alpha/\beta$.  
The minimization of the CV-error then occurs with respect to a single hyperparameter $\lambda$. 
This approximation is commonly referred to as ridge regression.\cite{scikit-learn} 

The minimization of the CV-error with respect to $\lambda$ 
 alone does not allow for an independent determination of $\alpha$ and $\beta$. 
Both sets of hyperparameters are required, however, to sample ECI vectors, $\vec{w}$, from the posterior distribution for downstream uncertainty quantification.
Additional considerations are required to specify $\beta$, with $\vec{\alpha}$ then equal to $\vec{\lambda}\beta$.
In the absence of truncation errors, $\beta$ represents the precision on the training data, with the variance on training data, $\sigma^2_t=1/\beta$, arising from numerical noise. 
This value can be estimated by performing convergence tests of the numerical calculations of the DFT energies. 
A truncation of the full cluster expansion, however, will in general incur additional errors when comparing energies predicted with a truncated cluster expansion $\vec{w}$ and the training data, $\vec{t}$ (Eq. \ref{eqn:cluster_expansion_plus_noise}). 
The value of $\beta$ that is then to be used in the likelihood function, Eq. \ref{eqn:likelihood_function}, should account for both the numerical noise on the training data as well as a truncation error. 
This is more difficult to assess. In general, it should be possible to determine any hyperparameters for Bayesian priors using cross-validation \cite{vehtari2002bayesian}, although the scaling of cross-validation makes this difficult as more hyperparameters are used. 

\subsubsection{Evidence approximation}
\label{sec:evidence_approximation}

The evidence approximation enables an independent estimate of the prior and likelihood hyperparameters, $\vec{\alpha}$ and $\beta$.  
It does this by maximizing $P(\vec{\alpha},\beta|\vec{t})$, the posterior distribution for $\vec{\alpha}$ and $\beta$, conditioned on the training data, $\vec{t}$. 
Using Bayes' theorem, this distribution can be written as \cite{Bishop2006}
\begin{equation}
    P(\vec{\alpha},\beta|\vec{t})\sim P(\vec{t}|\vec{\alpha},\beta)P(\vec{\alpha},\beta)
    \label{eqn:posterior_on_alpha_beta}
\end{equation}
where $P(\vec{t}|\vec{\alpha},\beta)$ is the marginalized likelihood
\begin{equation}
        P(\vec{t}|\vec{\alpha},\beta) = \int_{\vec{w}}P(\vec{t}|\vec{w},\beta)P(\vec{w}|\vec{\alpha})d\vec{w}.
        \label{eqn:marginalized_likelihood}
\end{equation}
Within the assumption that the prior probability, $P(\vec{\alpha},\beta)$, of Eq. \ref{eqn:posterior_on_alpha_beta} is nearly uniform, the values of $\vec{\alpha}$ and $\beta$ that maximize $P(\vec{\alpha},\beta|\vec{t})$ also maximize $P(\vec{t}|\vec{\alpha},\beta)$. 
When both the likelihood, $P(\vec{t}|\vec{w},\beta)$, and the prior, $P(\vec{w}|\vec{\alpha})$ are Gaussians, the marginalized likelihood $P(\vec{t}|\vec{\alpha},\beta)$ of Eq. \ref{eqn:marginalized_likelihood} can be integrated in closed form, yielding \cite{Bishop2006}  
\begin{align}
    \ln(P(\vec{t}|\vec{\alpha},\beta)) &= \ln \mathcal{N}(\vec{t}|\vec{0},\text{C}) \notag \\
    &= -\frac{1}{2} \left[ N\ln(2\pi) + \ln |\text{C}| + \vec{t}^{\mathsf{T}}\text{C}^{-1}\vec{t} \right] \label{eqn:log_evidence_approximation}
\end{align}
where the matrix C is given by
\begin{equation}
    \text{C} = \beta^{-1}\text{I} + X A^{-1}X^{\mathsf{T}}
\end{equation}
and $A$ is the diagonal matrix of precisions $\vec{\alpha}$. 
Within the evidence approximation, the hyperparameters $\beta$ and $\vec{\alpha}$ are set equal to the values that maximize Eq. \ref{eqn:log_evidence_approximation}.
When a common value is used for all the elements of $\vec{\alpha}$ the approximation is referred to as Bayesian ridge.\cite{scikit-learn,mackay1992bayesian}
In the relevance vector machine (RVM), each component of $\vec{\alpha}$ is allowed to vary independently when maximizing Eq. \ref{eqn:log_evidence_approximation}.\cite{tipping2001sparse,Bishop2006}
The RVM has been demonstrated as a useful approach for creating Bayesian cluster expansions. \cite{ALDEGUNDE2016}

\subsection{First-principles statistical mechanics calculations}

The Li$_x$Mg$_{1-x}$ and Li$_x$Al$_{1-x}$ alloys having the BCC parent crystal structure are used as model systems to explore the concepts described in the previous sections when constructing cluster expansions and their posterior distributions.
First-principles calculations of formation energies for the Li$_x$Mg$_{1-x}$ and Li$_x$Al$_{1-x}$ alloys were performed using various approximations to density functional theory (DFT) as implemented in the Vienna \textit{ab initio} Simulation Package (VASP) \cite{kresse1993ab,kresse1996efficiency,kresse1996efficient}.
Pseudopotentials based on the projector-augmented-wave method (PAW) \cite{blochl1994projector} with valence electronic configurations of $1s^22s^1$, $3s^2$, and $3s^23p^1$ were used for Li, Mg, and Al, respectively.
Calculations within the local density approximation (LDA)  used LDA pseudopotentials, while those using the Perdew Burke Ernzerhof (PBE) and the Strongly Constrained Appropriately Normalized (SCAN) approximations used the PBE pseudopotentials provided by VASP. 
However, for the SCAN calculations, the exchange-correlation functional was calculated using the METAGGA SCAN 
tag as implemented in VASP. 
All calculations used a plane-wave energy cut-off value of 650 eV (ENCUT) and a $\Gamma$-centered automatic k-point mesh generation scheme with a grid spacing of 2$\pi$/64 \AA{}$^{-1}$.
Geometric convergence was performed with a convergence threshold of 0.02 eV/\AA{} on forces, while electronic convergence was performed with a threshold of $10^{-5}$ eV on energies.
All calculations used Gaussian smearing with 0.02 eV smearing width while performing ionic relaxations. 
A final static run was performed on the converged geometry using the tetrahedron method with Bl\"ochl corrections \cite{blochl1994improved} to obtain accurate total energies. 

\section{Results}

To analyze the concepts and methods of Section \ref{sec:methods}, we construct cluster expansions and their posterior probability distributions for the BCC forms of Li$_x$Mg$_{1-x}$ and Li$_x$Al$_{1-x}$. 
Both alloys are currently of interest due to their beneficial role in moderating morphological evolution during lithium plating and stripping in all-solid-state Li-ion batteries.\cite{yang2019electron, krauskopf2019diffusion,liu2023aluminum,lee2020high}
The Li$_x$Mg$_{1-x}$ alloy forms a disordered solid solution on the BCC parent crystal structure above $x=0.3$, while the Li$_x$Al$_{1-x}$ alloy forms a variety of ordered phases on the BCC parent crystal structure above $x=0.5$.\cite{behara2024fundamental,behara2024role}

\begin{figure}[ht]
    \subfloat[]{\includegraphics[width=0.48\textwidth]{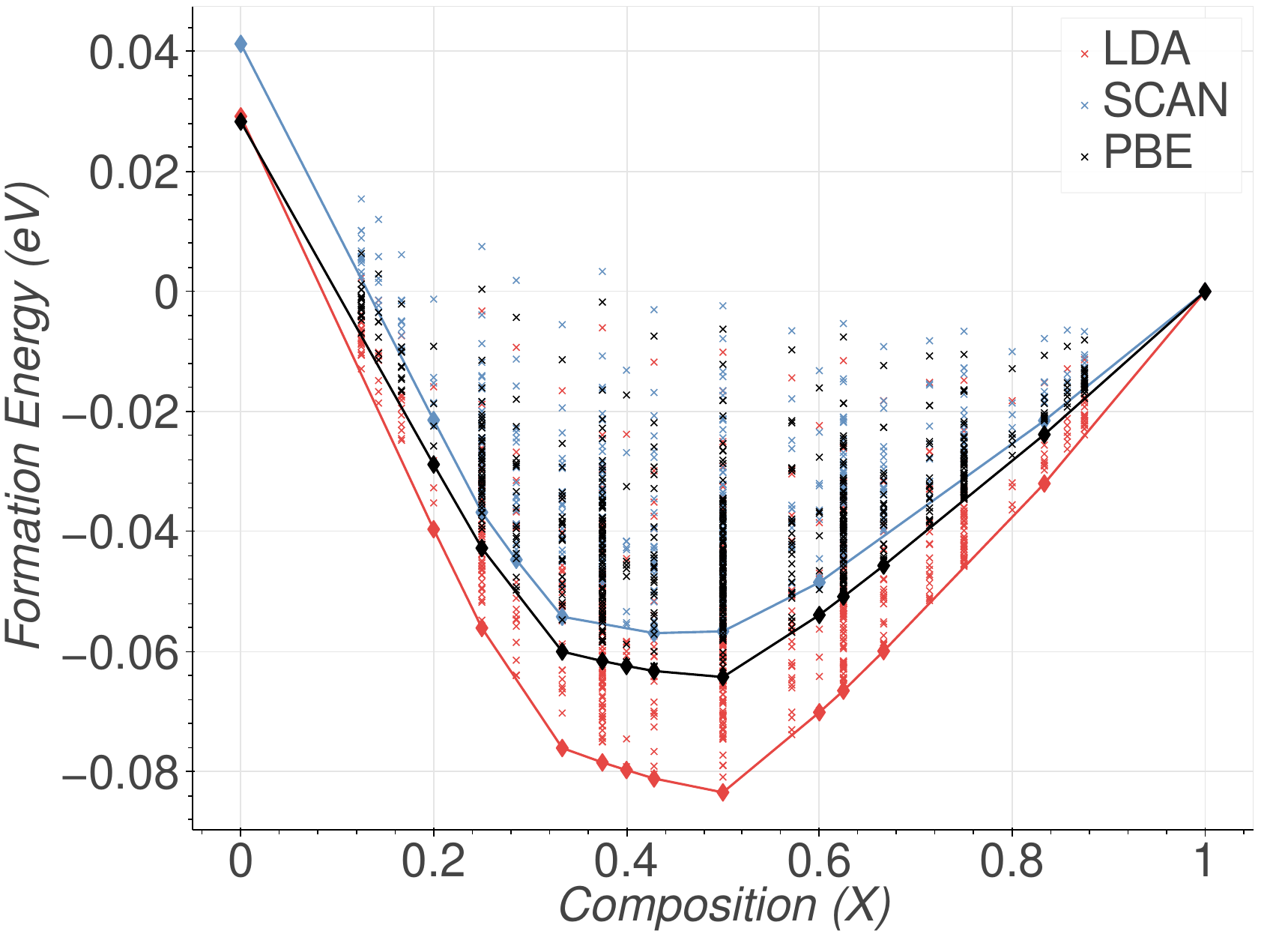}}\\
    \subfloat[]{\includegraphics[width=0.48\textwidth]{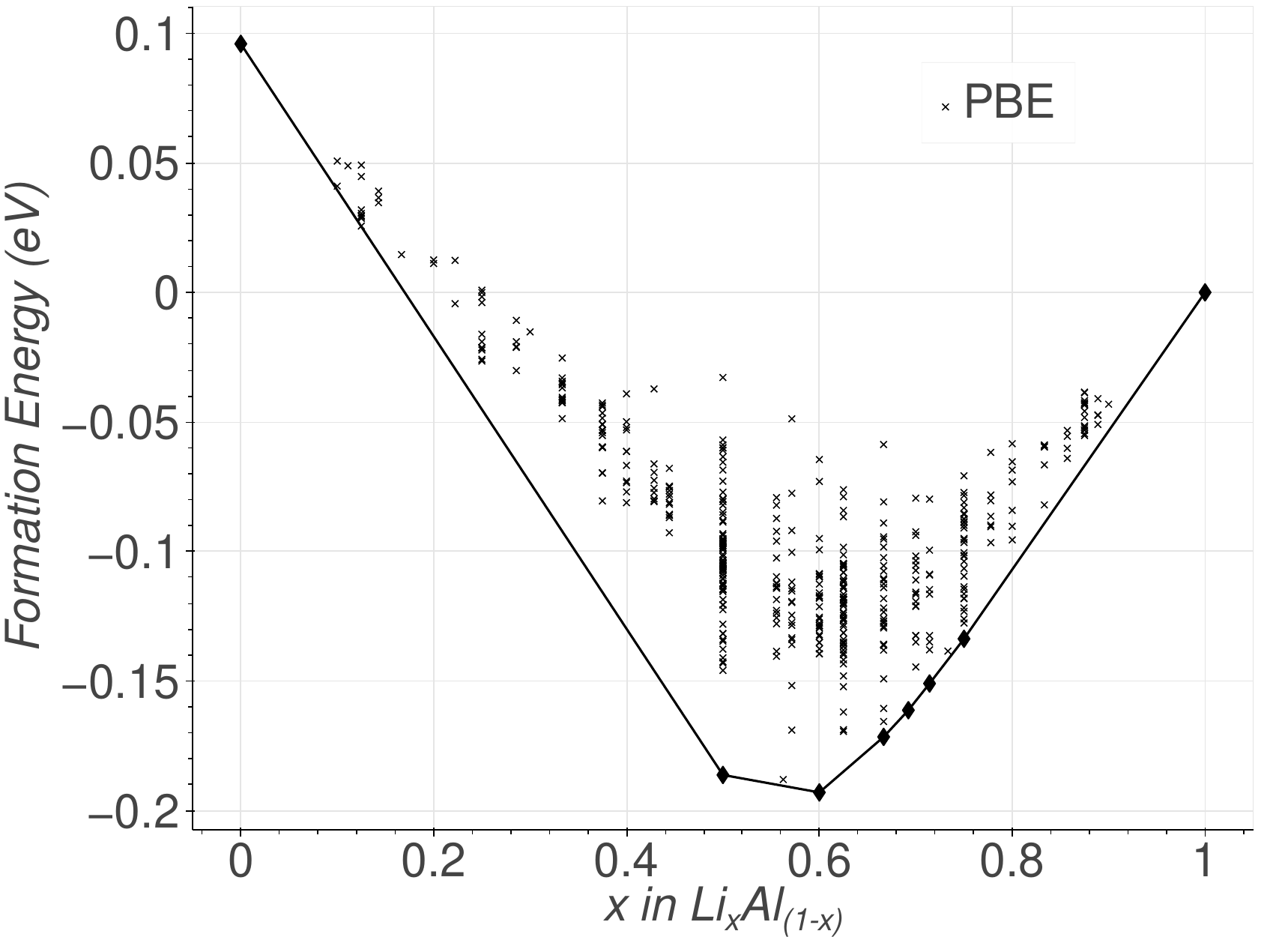}}
    \caption{(a) DFT formation energies and lower convex hulls for BCC Li$_x$Mg$_{1-x}$ calculated with LDA, PBE and SCAN. The energies of HCP Mg and BCC Li are used as reference states. (b) DFT formation energies for BCC Li$_x$Al$_{1-x}$ calculated with PBE. The energies of FCC Al and BCC Li are used as reference states.} 
    \label{fig:DFT_formation_energies}
\end{figure}

Crucial to constructing a posterior probability distribution for a cluster expansion surrogate model is training data. 
This takes the form of a set of formation energies for different arrangements of the alloy constituents on a parent crystal structure. 
For this study, the formation energies of 630 different Li-Mg configurations over the sites of the BCC crystal structure were calculated using three different DFT functionals: the Local Density Approximation (LDA) \cite{kohn1965self}, the Generalized Gradient Approximation from Perdew-Burke-Ernzerhof (PBE) \cite{perdew1996generalized} and the SCAN meta-Generalized Gradient Approximation functional \cite{sun2015strongly}. 
The calculated formation energies are shown in Figure \ref{fig:DFT_formation_energies}(a). 
The energies of BCC Li and HCP Mg were used as reference states when calculating the formation energies. 
Figure \ref{fig:DFT_formation_energies}(b) shows the formation energies of 444 Li-Al configurations over the sites of the BCC crystal structure as calculated with DFT-PBE. 
The Li-Al binary alloy is more complex than the Li-Mg alloy, as it forms a variety of intermetallic compounds that remain stable to high temperatures. 
Furthermore, many Al-rich configurations on the BCC parent crystal structure spontaneously relax along the Bain path to an ordering on the FCC parent crystal structure. 
These configurations cannot be used as training data for a cluster expansion of the BCC parent crystal structure. 
A total of 901 Li-Al configurations were considered, of which 457 relaxed away from the BCC parent structure, with most configurations relaxing to an ordering on the FCC parent crystal structure.
The configurations that maintained a BCC parent crystal structure were identified using a crystal structure mapping algorithm.\cite{thomas2021comparing} 

Two downstream quantities are used to assess the propagation of uncertainty in the training of cluster expansions: the configurational free energy and the electrochemical potential of the alloy with respect to a lithium reference anode. 
The electrochemical potential is proportional to the Li chemical potential in the alloy according to the Nernst equation.\cite{van2020rechargeable} 
These quantities are calculated with Monte Carlo simulations in which a particular cluster expansion model is used to calculate the energy of the crystal for different configurational microstates. 


\subsection{Hyper-parameter determination by minimizing the cross-validation score}

A common approach to determine the level of truncation of a cluster expansion and the prior precisions on the non-truncated expansion coefficients for a given training data set is to minimize a k-fold cross-validation error (Section \ref{sec:hyperparameters_through_cv_minimization}) \cite{van2002automating,zarkevich2004reliable,hart2005evolutionary,blum2005using}.  
Since the number of prior precisions, $\alpha_c$ (one for each ECI $w_c$) is generally large, it is common to set them all equal to the same value $\alpha$. 
This approximation is known as Ridge Regression when the likelihood and prior distribution are Gaussian.\cite{scikit-learn}
\begin{figure}
    \centering
\includegraphics[width=1\linewidth,height=1\textheight,keepaspectratio]{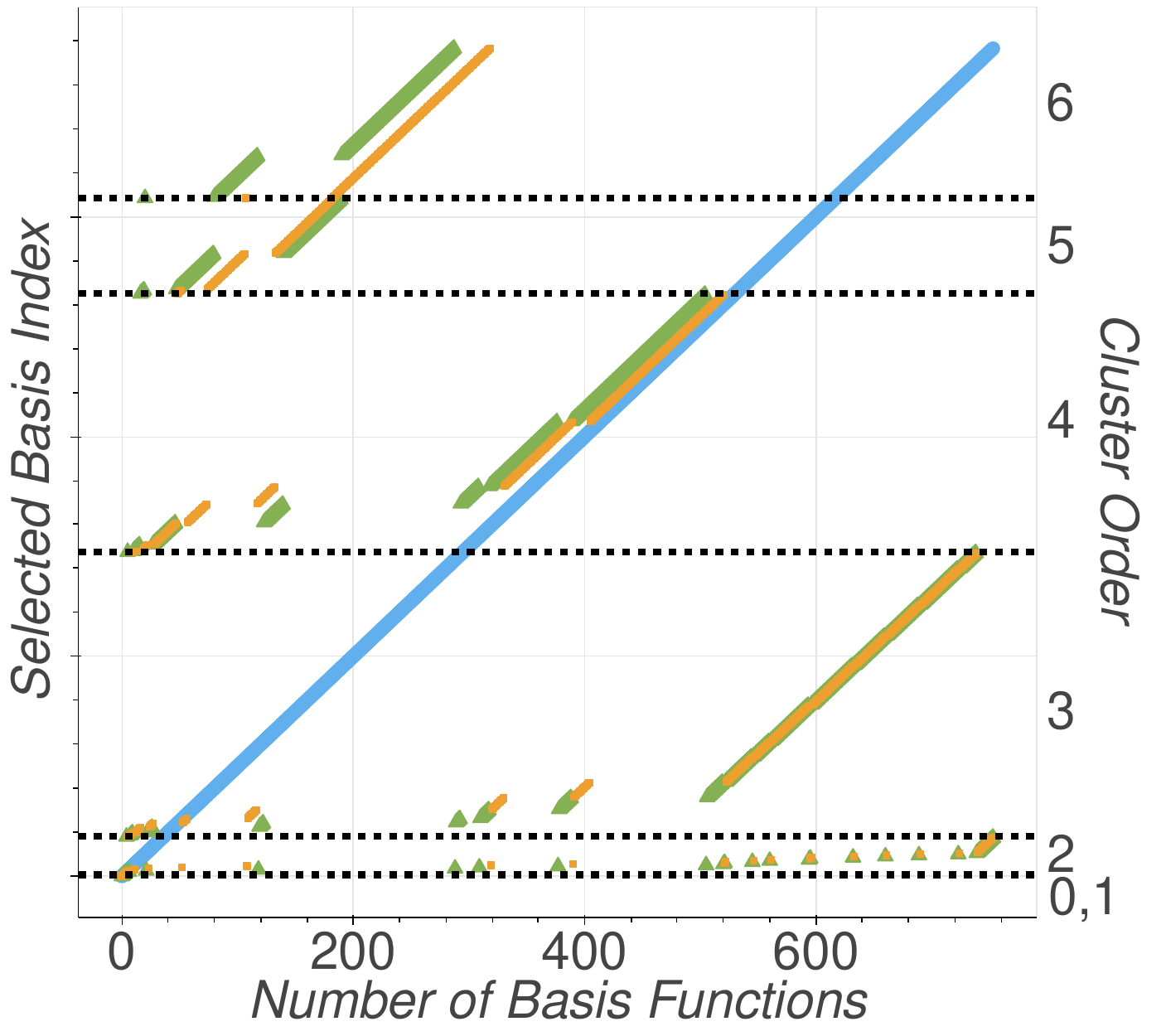}
    \caption{The order in which basis functions are incrementally added to a truncated cluster expansion for three sequences. The horizontal axis tracks the number of basis functions in the model, while the vertical axis stratifies the basis functions in bands based on the number of sites in their corresponding cluster. The lowest band collects two-site cluster basis functions, the second lowest band collects three-site cluster basis functions, etc. up through six-site cluster basis functions. Within each n-site cluster band, basis functions are sorted by the maximum pair distance within their corresponding cluster. Constant and point clusters are always added before any other higher body-order clusters. Cluster basis functions within each band are sorted by the largest pair distance within the cluster. Sequence I (blue circles) incrementally adds all n-site cluster basis functions before any (n+1)-site cluster basis functions are added. Sequence II (green triangles) incrementally adds cluster basis functions based on their maximum pair distance, prioritizing smaller clusters over larger clusters whenever there is a tie in the maximum pair distance. Sequence III (orange squares) is similar to Sequence II, but only adds (n+1)-site cluster basis functions if its maximum pair distance is strictly less than the maximum pair distance of the last n-site cluster basis function.}
    \label{fig:cluster_sorting}
\end{figure}
\begin{figure}[ht]
    \subfloat[]{\includegraphics[width=0.41\textwidth]{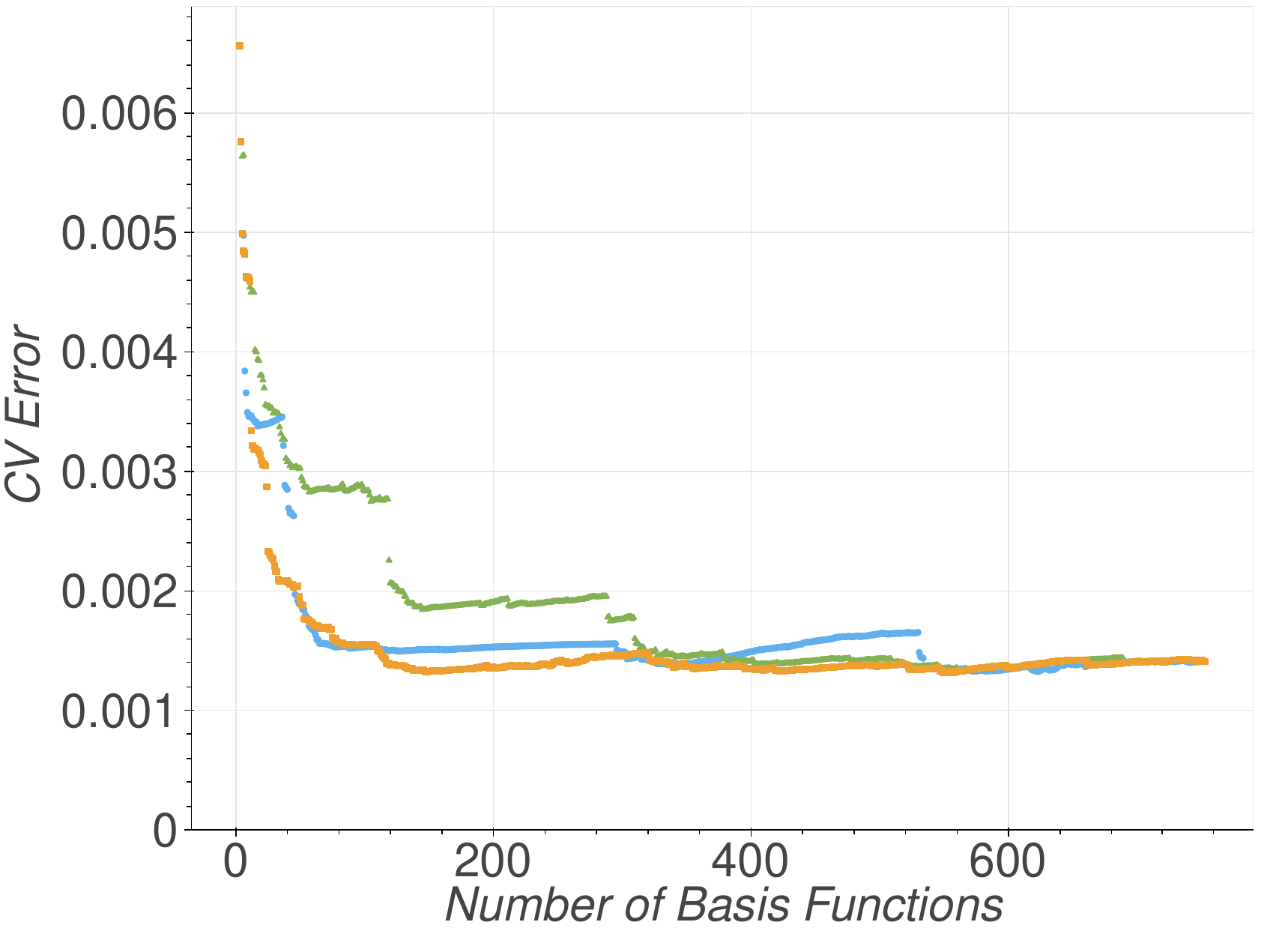}}\\
    \subfloat[]{\includegraphics[width=0.41\textwidth]{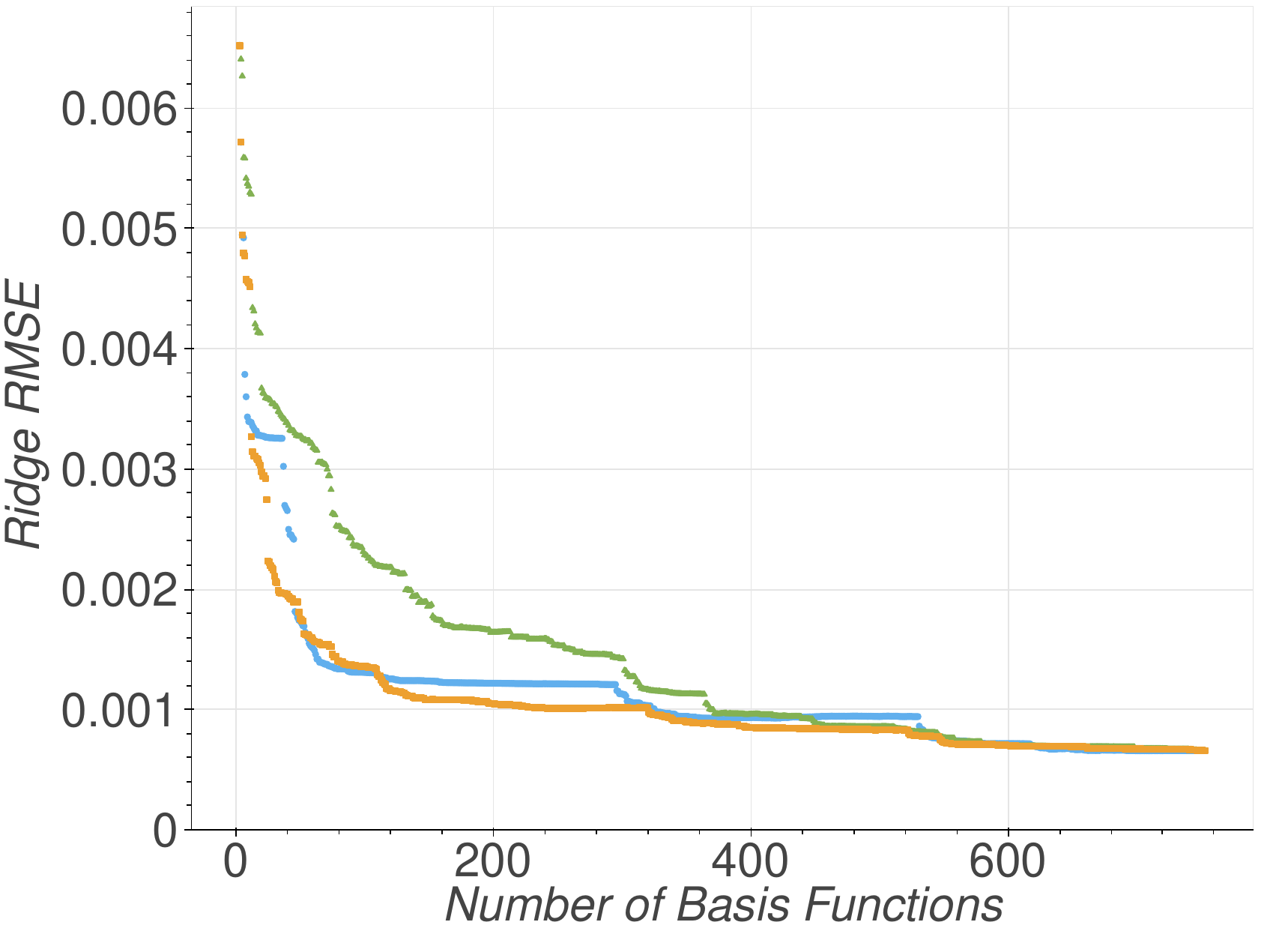}}\\
    \subfloat[]{\includegraphics[width=0.41\textwidth]{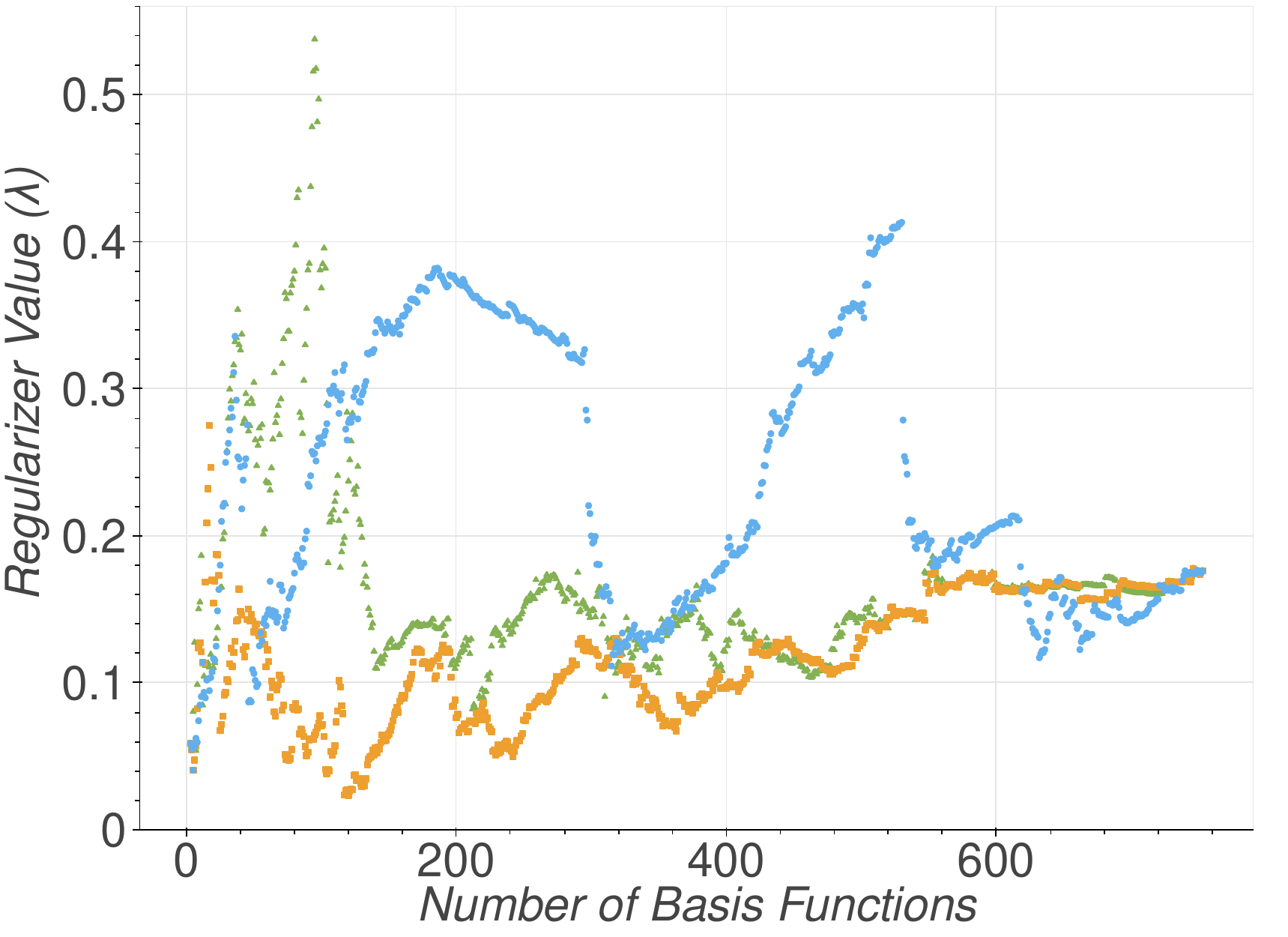}}
    \caption{(a) Cross-validation error (error units of eV) (b) RMSE and (c) Ridge regularizer $\lambda$ all vs model truncation according to cluster sorting sequences I, II, and III: blue circles, green triangles, and orange squares, respectively.} 
    \label{fig:CV_all}
\end{figure}


In studying cluster expansion parameterization using CV-optimization, we utilize the DFT-PBE training data set of the Li$_x$Mg$_{1-x}$ alloy, and the ``RidgeCV" implementation of leave-one-out cross-validation within scikit-learn \cite{scikit-learn}.
We explore three different sequences in which non-truncated cluster basis functions are incrementally added to the model. 
For each cluster expansion model, basis functions are drawn from a total of 754 candidate cluster basis functions that include 34 pair, 259 triplet, 235 quadruplet, 87 five-site, and 137 six-site clusters.
In sequence I, pair cluster basis functions of increasing pair distance are added incrementally before any triplets are added. 
Once all the pairs have been added, triplets are incrementally added in increasing order of the largest pair distance within the triplet cluster until all triplet clusters are included. 
This is repeated in a similar manner for the quadruplet, five-point, and six-point clusters.  
The order in which the basis functions are added in sequence I is shown by the blue circles in Figure \ref{fig:cluster_sorting}, which graphically shows that the pairs are incrementally added first, followed by the incremental addition of triplets, etc. 
In sequence II, clusters are added incrementally based on the largest pair distance within the cluster, starting from smallest to largest. 
When two clusters with a different number of sites have the same maximum pair distance, the one with fewer sites is added first. 
The order in which the basis is grown in sequence II is shown by the green triangles in Figure \ref{fig:cluster_sorting}. 
Sequence III is similar to sequence II except that only higher body clusters are included if their maximum pair length is strictly less than (but not equal to) the maximum pair length of the lower body order cluster. 
Sequence III is shown by the orange diamonds in Figure \ref{fig:cluster_sorting}. 
Sequence I prioritizes lower body order clusters while sequences II and III prioritize small and compact clusters, independent of the number of sites. 
Sequence II has the most aggressive incorporation of higher body clusters. 

Figure \ref{fig:CV_all}(a) shows the leave-one-out ($k=n$) cross-validation score for the three different sequences as a function of the number of basis functions. 
At each level of truncation (i.e. for each number of basis functions), a common hyperparameter $\lambda=\alpha/\beta$ was optimized to minimize the cross-validation score as described in Section \ref{sec:hyperparameters_through_cv_minimization}. 
The CV-error for Sequence III (orange) decreases most rapidly and achieves its minimum CV-error after the inclusion of approximately 150 basis functions. 
Sequence I shows the second fastest decrease in the CV-score. 
Figure \ref{fig:CV_all}(a) shows that the sequence I (blue) CV-error plateaus after the 9$^{\text{th}}$ nearest neighbor pair is added and only drops significantly upon addition of triplet cluster basis functions. 
The CV-error of sequence II, which favors compact clusters independent of the number of sites, decreases much more slowly than sequences I and III and only achieves similar optimal CV-errors after the basis set size exceeds 400. 
Figure \ref{fig:CV_all}(b) shows the root mean square error (RMSE) between cluster expansion predictions and the training data as a function of the number of basis functions for each truncation sequence. 
The RMSE, while slightly lower than the CV-error, exhibits similar trends with the number of basis functions, but does not exhibit a minimum or level off.  
Note also that the RMSE never becomes equal to zero, even when the number of basis functions exceeds the number of training data. 
This is because the prior distribution on the ECI regularizes each cluster expansion model to minimize the CV-error for a given level of truncation. 

Figure \ref{fig:CV_all}(c) shows the optimal hyperparameter $\lambda=\alpha/\beta$ as a function of basis size for the three sequences. 
Since the same hyperparameter, $\lambda$, is used for all non-truncated cluster basis functions, its value varies strongly with basis size. 
For example, in sequence II (green triangles), where 5 and 6 site clusters are added early on in the sequence, $\lambda$ increases rapidly. 
It decreases abruptly when larger pair and triplet clusters are added. 
Sequence III (orange squares), which has the fastest decrease in the CV-error and incorporates higher body clusters more gently than sequence II, has the most uniform variation of $\lambda$ with basis size. 
Sequence I (blue circles) tends to drop when traversing from one body order to another. 
However, $\lambda$ tends to increase when basis functions with the same body order but larger radius are incrementally added. 
A possible explanation for this trend is that longer-distance interactions need increased regularization to prevent overfitting. 

\subsection{Hyperparameter determination using the evidence approximation}
As reviewed in Section \ref{sec:evidence_approximation}, the evidence approximation provides a criterion for independently selecting the hyperparameters, $\vec{\alpha}$ and $\beta$, of the posterior distribution of a cluster expansion. 
Within the evidence approximation, the hyperparameters $\vec{\alpha}$ and $\beta$ are chosen to maximize the log evidence, Eq. \ref{eqn:log_evidence_approximation}.
In this section, we use only a single precision $\alpha$  (i.e. $\alpha_c$=$\alpha$ for all $c$) within the evidence approximation. 
Model selection using a common single precision for all ECI, $w_c$, is done using the ``BayesianRidge" (BR) function from scikit-learn.\cite{scikit-learn} 
In Section \ref{sec:RVM}, we analyze the evidence approximation when all precisions $\alpha_c$ are independently optimized within $\vec{\alpha}$; this is the Relevance Vector Machine (RVM). 

\begin{figure}[!h]
    \subfloat[]{\includegraphics[width=0.41\textwidth]{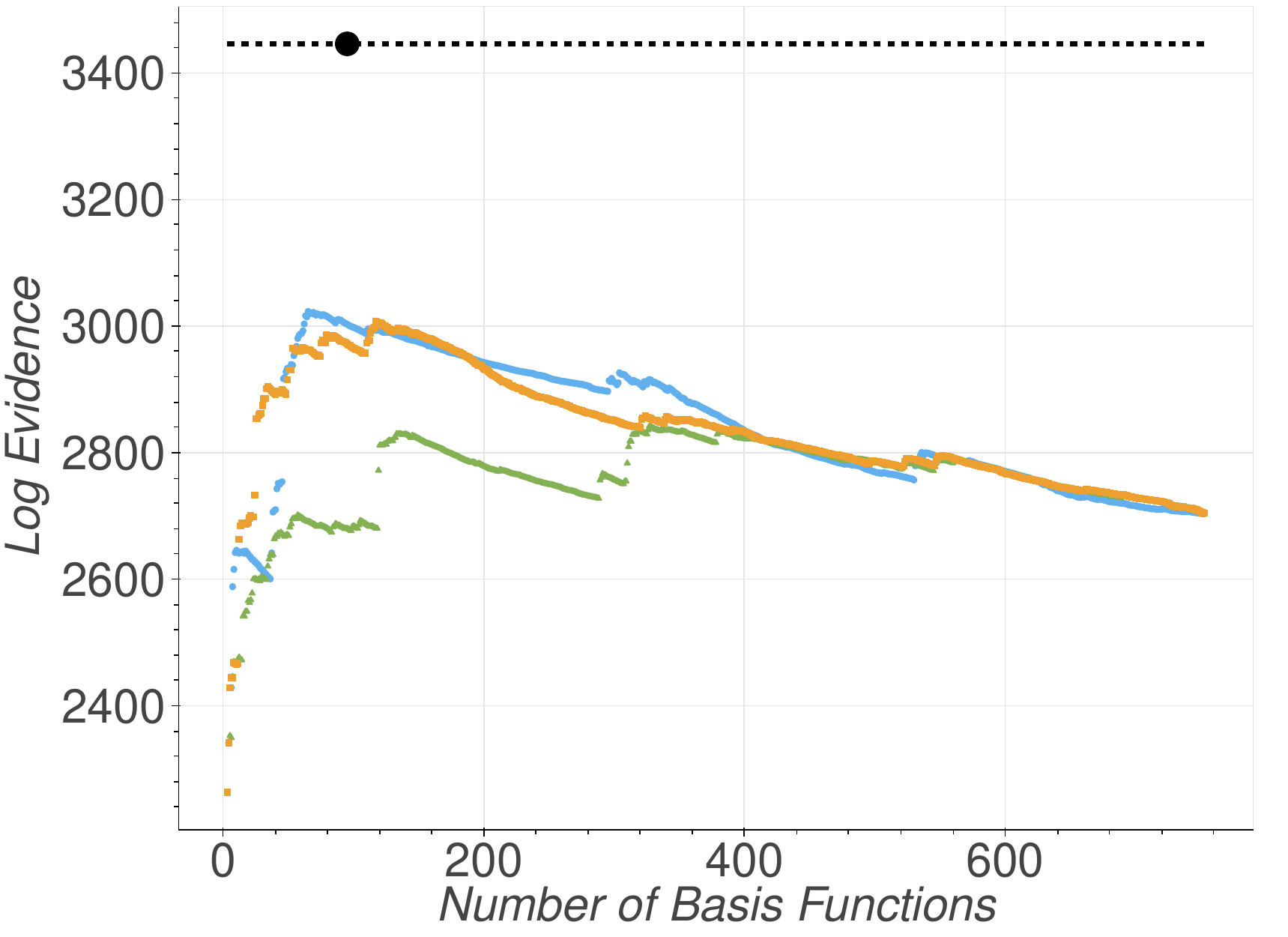}}\\
    \subfloat[]{\includegraphics[width=0.41\textwidth]{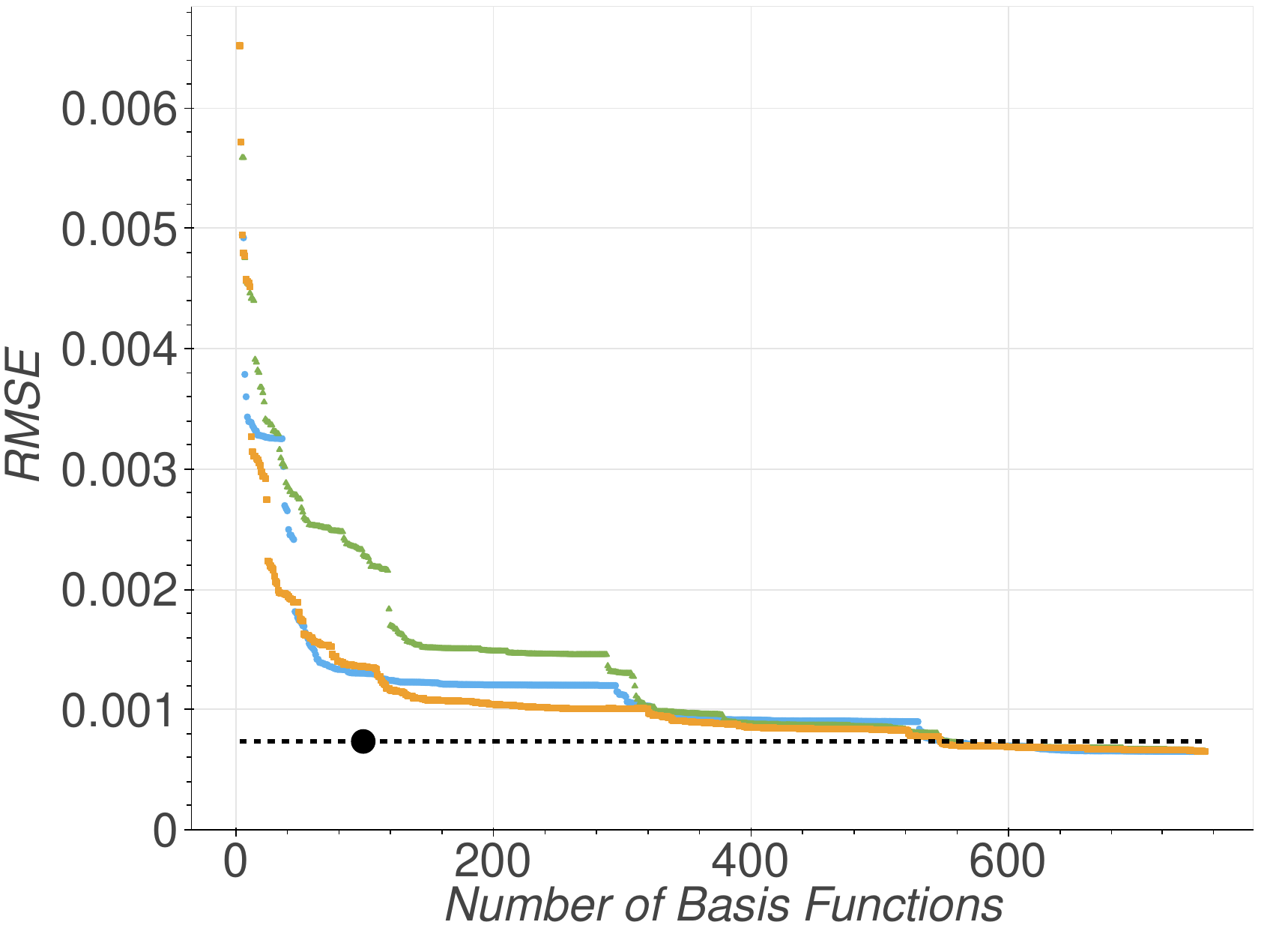}}\\
    \subfloat[]{\includegraphics[width=0.41\textwidth]{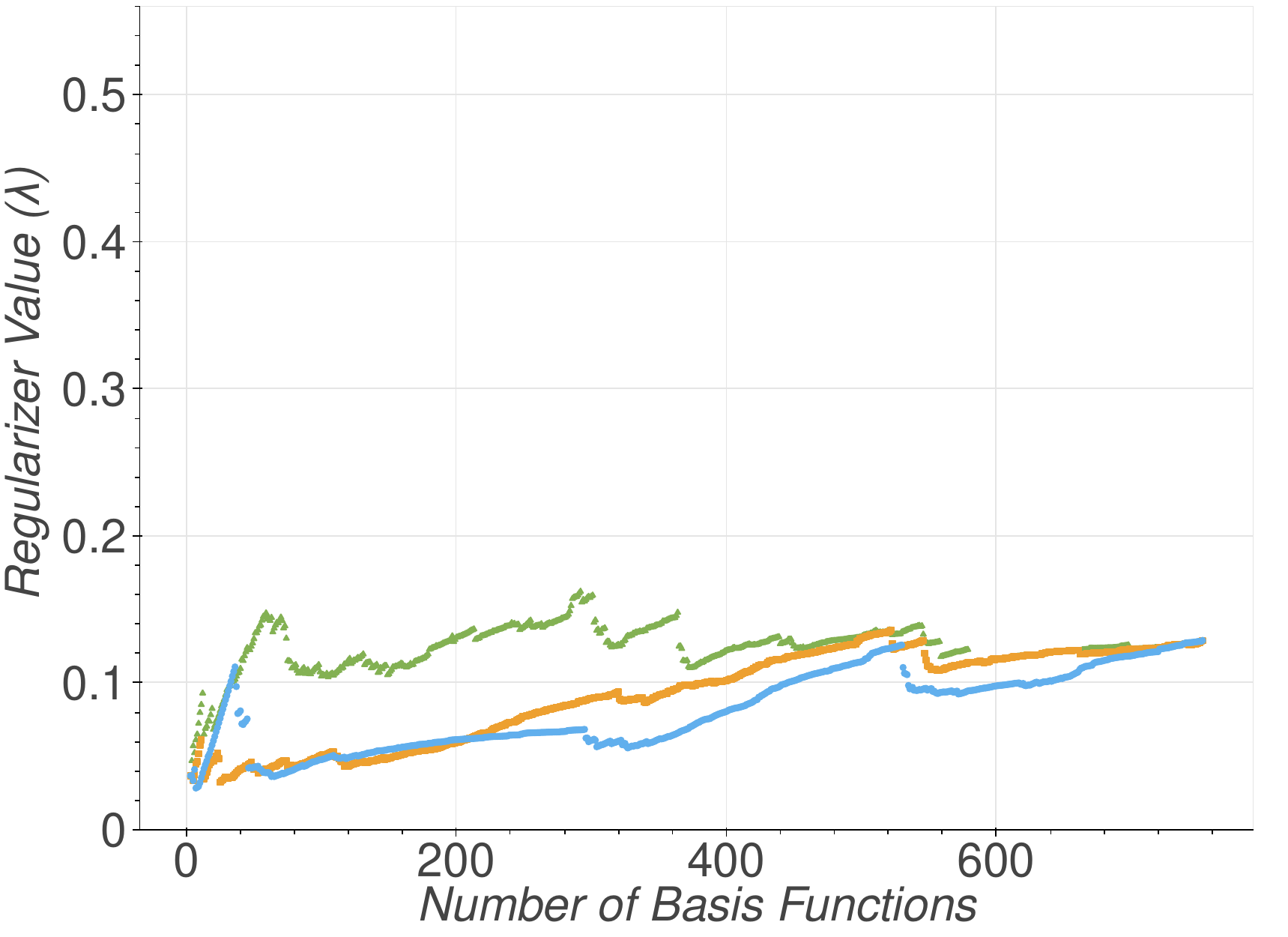}}
    \caption{(a) Evidence, (b)  model RMSE with respect to training data, (c) BR-implied regularizer vs model complexity. The three curves (blue circles, green triangles, and orange squares) correspond to sequences I, II, and III of cluster addition, respectively. Each point is the result of the MAP model from the BR approximation. The black point corresponds to the RVM MAP model, which selects its own basis from the full set of available basis functions. The horizontal dashed line through the RVM point is included to aid visual comparison between the RVM and BR approximation.} 
    \label{fig:evidence_vs_truncation}
\end{figure}

The evidence approximation is an alternative to the CV-optimization approach described in Section \ref{sec:hyperparameters_through_cv_minimization} and analyzed in the previous section. 
To compare the evidence approximation to the cross-validation minimization approach, we again use the DFT-PBE training data set of the Li$_x$Mg$_{1-x}$ alloy. 
We also use the same three sequences in which non-truncated basis functions are incrementally added to the model as described in the previous section. 
Comparison of the three cluster addition schemes is done by gradually exposing the BR method to an increasing number of basis functions, according to each of the three cluster addition sequences. 
Each time a basis function is added, the BR method is re-applied to each new basis set, selecting a model precision $\beta$ and a single ECI precision $\alpha$ that maximizes the log evidence of Eq. \ref{eqn:log_evidence_approximation}. 
Applying the BR method to the three sequences produces three curves for the log evidence (Eq. \ref{eqn:log_evidence_approximation}), the RMSE and $\lambda=\alpha/\beta$ as shown in Figures \ref{fig:evidence_vs_truncation}(a) and (b) 

Figure \ref{fig:evidence_vs_truncation}(a) shows that the evidence initially increases with the addition of basis functions for each truncation sequence. 
However, all BR methods show peaks in the log evidence, followed by a gradual decline as more basis functions are added. 
Similar to the cross-validation error of Figure \ref{fig:CV_all}(a), the model evidence of Figure \ref{fig:evidence_vs_truncation}(a) accounts for both model error and model complexity. 
The initial rise in log evidence is similar to the initial decrease in CV error, where additional basis functions reduce model error. 
However, the BR approximation cannot explicitly remove basis function dimensions once they are available. 
As additional basis functions are added beyond the peak in log evidence, the BR approximation must waste probability density on these new dimensions, reducing the model evidence. 
This is similar to the gradual increase in CV error in Figure \ref{fig:CV_all}(a) as additional basis functions are added.  
Beyond the point of minimal CV error, the error gradually rises as models become too complex and generalize poorly to data that they have not been trained on. 
Fundamentally, both evidence and cross-validation are balancing model error with model complexity.

Figure \ref{fig:evidence_vs_truncation}(b) shows the root-mean-squared error (RMSE) (between the cluster expansion predictions and the training data) of each truncation sequence versus the number of available basis functions. 
The RMSE decreases in a step-like manner, where each drop coincides with a step-increase in log evidence.
The RMSE values and trends of Figure \ref{fig:evidence_vs_truncation}(b) are similar to those of Figure \ref{fig:CV_all}(b) from the analogous cross-validation procedure. 

While cross-validation selects a value for $\lambda = \alpha/\beta$ to prevent overfitting and to regularize under-determined solutions, the BR approximation independently selects the model precision $\beta$ and prior precision $\alpha$. 
Therefore, it is possible to compare the cross-validation regularizers $\lambda$ from Figure \ref{fig:CV_all}(c) to the implied regularizers from the BR evidence approximation $\lambda =  \alpha/\beta$ shown in Figure \ref{fig:evidence_vs_truncation}(c). 
Both methods show similar orders of magnitude and similar spikes in regularizer values at similar basis truncation levels. 
However, the fact that BR regularizers do not exactly match the cross-validation regularizers is not surprising, considering that cross-validation and the BR determine hyperparameters by optimizing different quantities. 


\begin{figure}
    \centering
\includegraphics[width=1\linewidth,height=1\textheight,keepaspectratio]{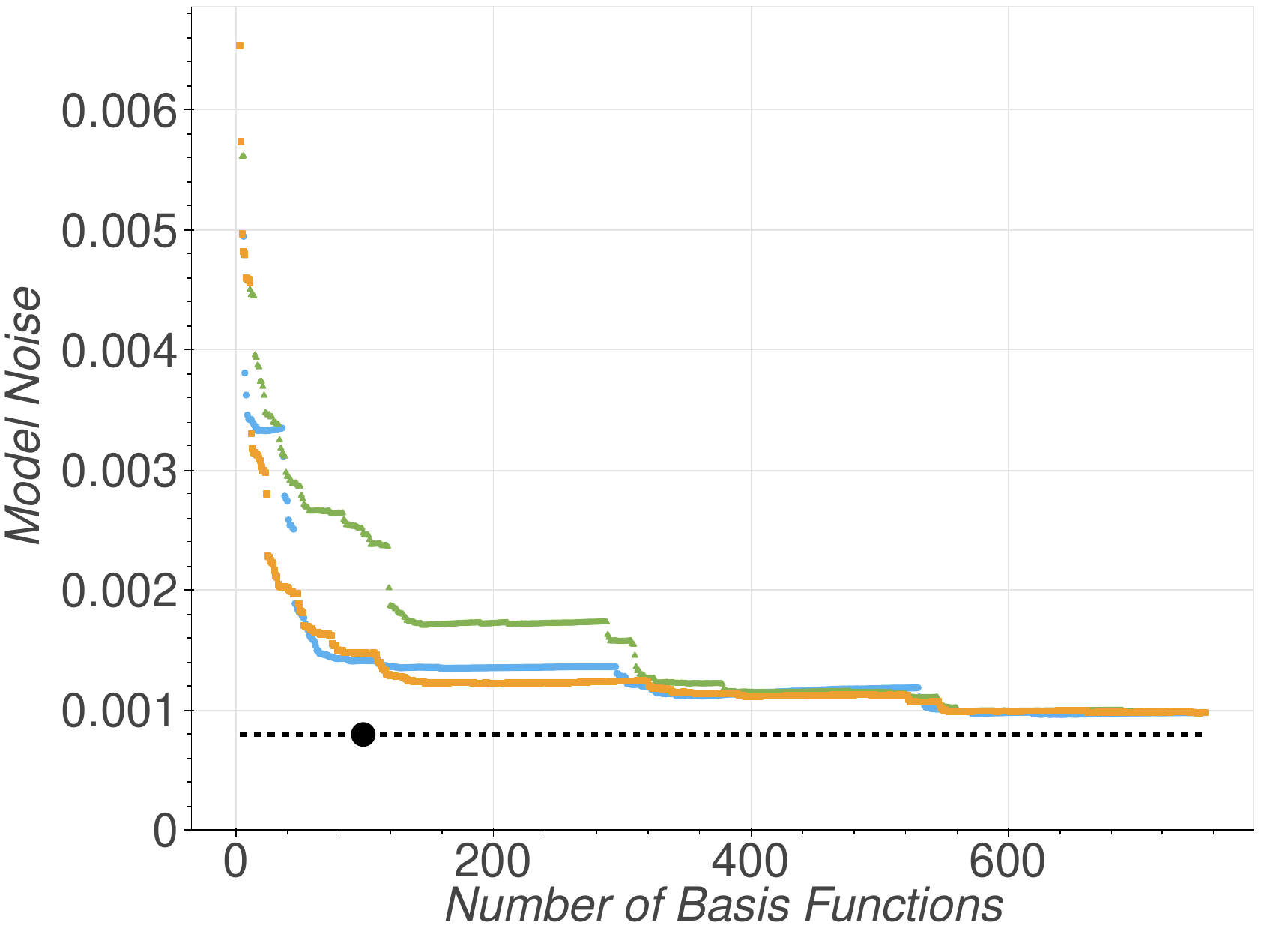}
    \caption{Model noise $(\sigma_t = 1/\sqrt{\beta})$ For the RVM MAP solution (black point) and the BR approximation for cluster sequences I, II, III as blue circles, green triangles, and orange squares, respectively.}
    \label{fig:model_noise_vs_truncation}
\end{figure}

Figure \ref{fig:model_noise_vs_truncation} shows the model noise ($1/\sqrt{\beta}$) as the BR method is exposed to additional basis functions according to the different truncation sequences. 
The cluster expansion is an exact model when all cluster basis functions are included (an infinite number for a crystal in the thermodynamic limit). 
Before truncation, the model noise is equal to the numerical noise on the training data, $\sigma_t$, which in this case arises from numerical noise of the DFT calculations. 
Truncation from an infinite to a finite basis will result in an additional error that then contributes to the model precision $\beta$ appearing in the likelihood function of Eq. \ref{eqn:likelihood_function}. 
Hence, as additional basis functions are added, the model noise $1/\sqrt{\beta}$ should asymptotically approach the numerical noise of the DFT calculations used to generate the training data. 
Figure \ref{fig:model_noise_vs_truncation} shows this decaying trend in model noise, regardless of truncation scheme. 
All methods converge to a noise of about 1meV per atom. 
The step-like drops in model noise also coincide with the step-like increases in log evidence for each algorithm. 

\subsection{Hyperparameter determination using the Relevance Vector Machine}
\label{sec:RVM}
Unlike the BR approximation, the relevance vector machine (RVM) selects individual values for each ECI precision $\alpha_c$. 
The RVM can thereby select an optimal truncation scheme and ``turn off" individual ECIs, $w_c$, by setting their ECI precision, $\alpha_c$, to infinity. 
This makes it possible to expose the RVM to the full basis set, and there is no need for a predefined truncation scheme. 
As a result, only one point (in black) is shown for the RVM model in Figures \ref{fig:evidence_vs_truncation}(a,b) and \ref{fig:model_noise_vs_truncation}.  
A horizontal dashed line is traced through the RVM point to facilitate visual comparison between the RVM and BR methods. 
All RVM models are selected according to our implementation of the RVM algorithm provided by Bishop \cite{Bishop2006}, which comes from the original work by Tipping \cite{tipping2001sparse}. 

The RVM log evidence is significantly higher when compared to the log evidence of BR methods in Figure \ref{fig:evidence_vs_truncation}(a). 
The difference between the RVM and BR models (black point vs colored curves) arises because RVM has many more hyperparameters that can be tuned independently. This grants the RVM more flexibility in distributing probability density in the large vector space of ECIs. 
Because the model evidence $P(\vec{t}|\vec{\alpha},\beta)$ of Eq. \ref{eqn:marginalized_likelihood} is the integral of the product of the likelihood function and the prior with respect to $\vec{w}$, it is maximized when there is high overlap between the probability densities of the likelihood and the prior. 
The RVM allocates probability density across each ECI axis independently to maximize the evidence, while the BR is constrained to use the same hyperparameter $\alpha$ for all ECI dimensions accessible by $\vec{w}$. 
If a particular basis function provides minimal gain in evidence, the RVM will set the corresponding precision to infinity. 
This forces the corresponding ECI to zero and allocates zero probability density across that basis function axis in ECI space, removing it from the fit. 
Figures \ref{fig:evidence_vs_truncation}(a,b) and \ref{fig:model_noise_vs_truncation} show this sparsification effect, where the optimal RVM solution selects approximately 100 cluster bases, setting the rest of the ECIs to zero.

Figure \ref{fig:evidence_vs_truncation}(b) shows that the RVM RMSE is slightly larger than the lowest RMSE obtained with BR. 
However, the BR models with the minimum RMSE have a larger number of ECI (i.e. basis functions) than the RVM model and do not necessarily generalize well to new data that they have not previously been trained on. 
In contrast to the RMSE, Figure \ref{fig:model_noise_vs_truncation} shows that the RVM model noise ($\sigma_t = 1/\sqrt{\beta}$) is lower than that produced by any BR model. 

\begin{figure}
    \centering
    \includegraphics[width=1.0\linewidth]{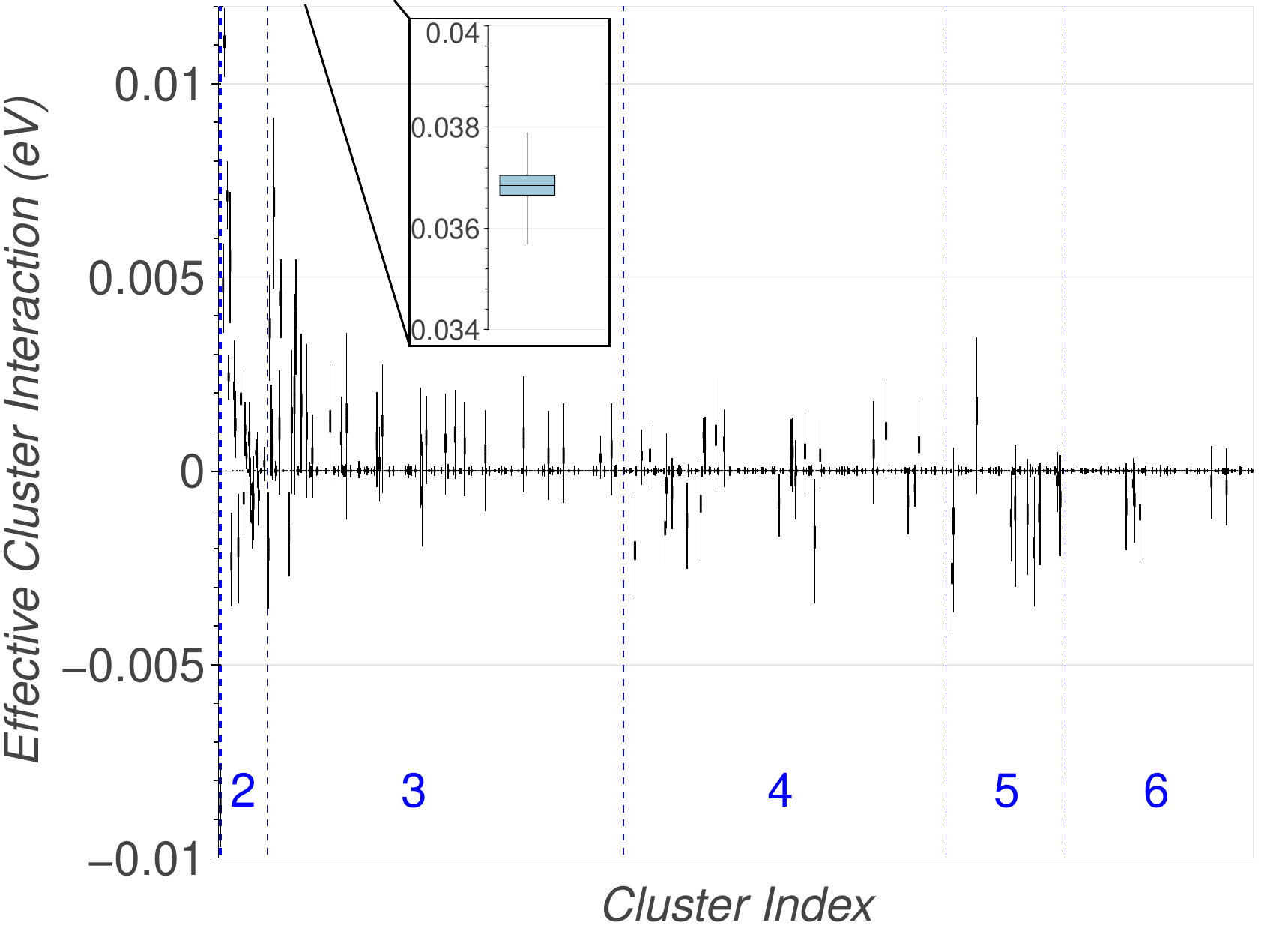}
    \caption{Boxplots of ECI uncertainty for all 754 ECIs in the RVM model, for PBE LiMg. Only 95 ECIs have magnitudes greater than 10$^{-4}$ eV.  Body orders are divided by vertical blue dashed lines, up to six-body interactions. Displayed ECI values are not normalized by symmetric multiplicity. The y-axis has been truncated to show uncertainties. Not shown is the constant term with a median value of -0.04589 eV, and first and third quartile values of -0.04595 and -0.04583 eV, respectively.}
    \label{fig:PBE_LiMg_RVM}
\end{figure}

Figure \ref{fig:PBE_LiMg_RVM}(a) shows the RVM-determined posterior ECI mean values and corresponding standard deviations in the full 754 cluster basis space. 
In the posterior distribution, only 95 of the 754 available clusters have ECIs with magnitudes greater than $10^{-4}$ eV (the basis functions are sorted according to Sequence I, i.e. low to high body order, and short to long pair distance within a body order). 
The basis functions with a negligible mean (i.e. $< 10^{-4}$ eV) span all cluster sizes between two-site and six-site clusters. 
Although the majority of the available basis functions correspond to four, five and six-site clusters, the majority of RVM-selected basis functions correspond to two-site and three-site clusters. 
Among the two and three-body interactions, there is a general preference for shorter-range interactions.
This result is consistent with physical intuition in that important contributions to a cluster expansion tend to involve few-body, short-range interactions.

\subsection{DFT vs Cluster Expansion Uncertainty: LiMg}
\label{sec:dft_uncertainty}


In this section, the sensitivity of macroscopic thermodynamic properties (free energies and voltage curves) on DFT approximation is assessed. 
The Li$_x$Mg$_{1-x}$ alloy on the BCC parent crystal structure is used as a model.
As is evident in Figure \ref{fig:DFT_formation_energies}, there are notable differences in the formation energies in the Li$_x$Mg$_{1-x}$ alloy system when calculated with different approximations to DFT. 
LDA predicts more negative mixing energies than PBE and SCAN. 
Furthermore, the ground state at $x=1/2$ (which has the B2 ordering on BCC \cite{kolli2020discovering}) is predicted to be more stable relative to the other ordered phases with the LDA functional than with the PBE and SCAN functionals. 
While LDA and PBE predict a similar difference in energy between HCP Mg (which has a zero formation energy at $x=0$) and BCC Mg, that predicted by SCAN is substantially larger. 
All three functionals predict a large number of Li-Mg orderings that are on or very close to the convex hull. 
This suggests that there is no strong preference for any particular long-range order among Li and Mg on the BCC parent crystal and that ordered compounds on the convex hull will disorder at relatively low temperatures. 

We used the RVM method to determine hyperparameters $\vec{\alpha}$ and $\beta$ for the posterior distributions of cluster expansion models trained to the three sets of formation energies as calculated with LDA, PBE, and SCAN. 
The resulting posterior distribution quantifies the cluster expansion model uncertainty. 
Sampling ECI vectors from this posterior distribution and propagating them through thermodynamic Monte Carlo calculations produces a probabilistic free energy and voltage curve for Li$_x$Mg$_{1-x}$. 
This posterior distribution quantifies the uncertainty in the cluster expansion model for a given DFT training data set. 

\begin{figure}[!h]
    \subfloat[]{\includegraphics[width=0.45\textwidth]{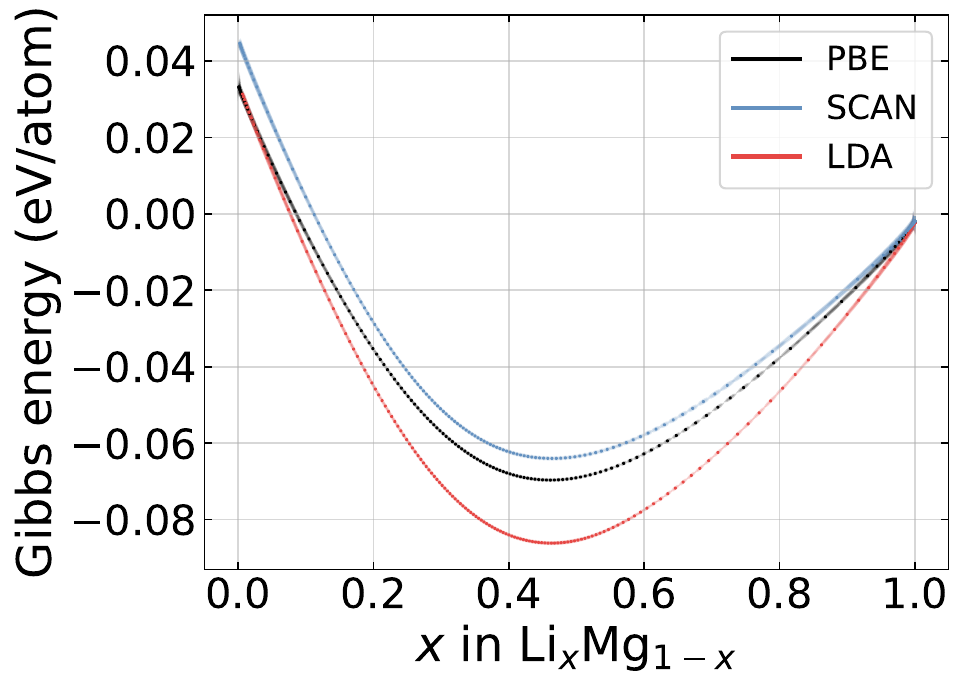}}\\
    \subfloat[]{\includegraphics[width=0.45\textwidth]{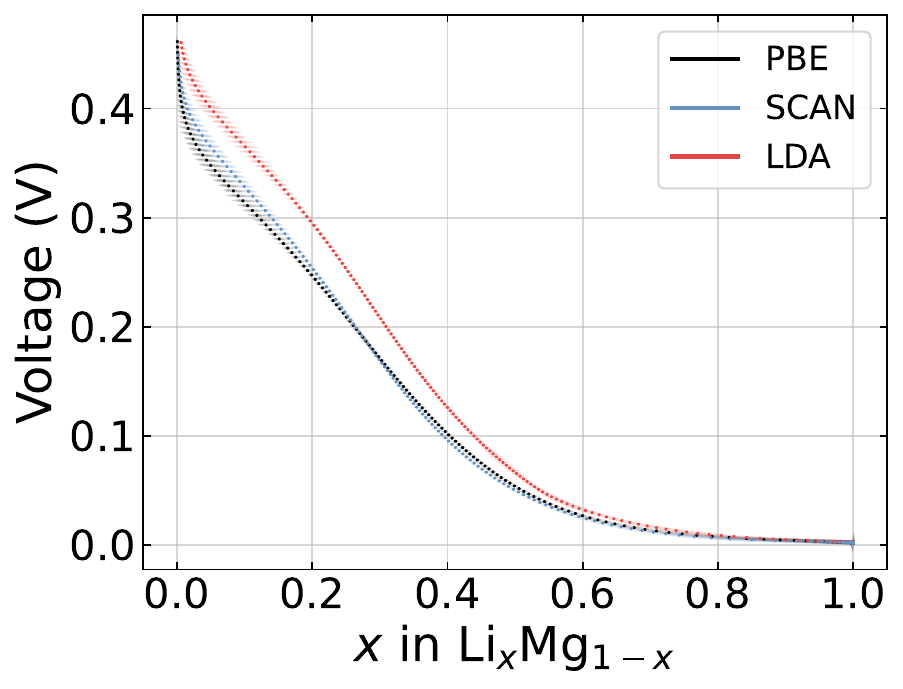}}
    \caption{Li--Mg Gibbs free energy and voltage. Propagated uncertainties are depicted as 95\% confidence ellipses around each data point.} 
    \label{fig:LiMg_voltage}
\end{figure}

A total of 100 ECI vector samples, $\vec{w}$, were drawn from each of the three posterior distributions (i.e. using LDA, PBE and SCAN training data in the likelihood function, Eq. \ref{eqn:likelihood_function}). 
Each sample was propagated through semi-grand canonical Monte Carlo calculations to produce 100 Gibbs free energy curves, and 100 voltage curves for each DFT data set. 
Figure \ref{fig:LiMg_voltage}(a) shows the average Gibbs free energy calculated in this manner at 300 K for the three approximations to DFT along with 95\% elliptical confidence intervals. 
Figure \ref{fig:LiMg_voltage}(b) uses the same form of 95\% confidence intervals but for composition and voltage. 
In both cases, the variability across different DFT functionals is larger than cluster expansion uncertainty for a given functional. 
Note that Figure \ref{fig:LiMg_voltage} only contains points and confidence ellipses; there are no lines or splines connecting points. 
The thermodynamic Monte Carlo simulations were performed in the semi-grand canonical ensemble, where the exchange chemical potential (i.e. $\tilde{\mu}=\mu_{Li}-\mu_{Mg}$, which is equal to the slope of the free energy) \cite{puchala2025casmMC} is imposed while the average composition, $x$, is calculated. 
Since the slope of the Gibbs free energy is equal to the imposed exchange chemical potential ($\tilde{\mu}=\mu_{Li}-\mu_{Mg}$), while the composition is allowed to fluctuate in the grand canonical Monte Carlo,\cite{puchala2025casmMC} the confidence ellipses on the Gibbs free energies are aligned along the free energy curves.
There is little uncertainty in the predicted voltage because the voltage itself is related to the lithium chemical potential through the Nernst equation.\cite{van2020rechargeable} 
Instead, as is evident in Figure \ref{fig:LiMg_voltage}(b), the confidence ellipses are predominantly horizontal.

\subsection{Enforcing Ground States}
\label{sec:enforcing_groundstates}


The BCC Li$_x$Mg$_{1-x}$ alloy has many ordered phases with formation energies that reside on the convex hull. 
These ground state orderings, however, are weakly stable and disorder below room temperature. 
The qualitative nature of the Li$_x$Mg$_{1-x}$ ground state orderings therefore has little effect on predicted room temperature thermodynamic properties. 
In contrast, the Li$_x$Al$_{1-x}$ alloy has ground state orderings that remain stable well above room temperature and influence the topology of room temperature thermodynamic quantities. 
The intermetallic compounds with compositions $x > 0.5$ in the Li$_x$Al$_{1-x}$ alloy system are superlattice orderings on the BCC parent crystal structure.\cite{behara2024fundamental} 
Each intermetallic compound appears as a single phase region in the Li-Al temperature versus composition phase diagram.\cite{mcalister1982li,sigli1986calculation} 
Furthermore, each intermetallic compound produces a step in the voltage profile when the alloy is used as an electrode in a Li-ion battery with respect to a lithium metal reference anode.\cite{wen1979thermodynamic,ghavidel2019electrochemical,van2020rechargeable} 
A cluster expansion for the Li$_x$Al$_{1-x}$ alloy that predicts incorrect ground states will result in qualitatively incorrect downstream predictions of thermodynamic properties. 


The approaches of the previous sections do not guarantee that cluster expansion models sampled from a posterior distribution predict the correct ground states or even a consistent set of ground states. 
For example, an RVM model was constructed using the formation energies of 444 Li-Al orderings on the BCC parent crystal structure and 505 candidate cluster basis functions. 
The RVM model retains 97 cluster basis functions (with corresponding non-zero ECI) and its maximum {\it a posteriori} (MAP) cluster expansion, $\vec{w}_{MAP}$, has an RMSE of 3.3 meV per atom between training and predicted formation energies. 
While this RMSE error is small, the RVM MAP cluster expansion does not replicate the DFT-PBE ground state set. 
Furthermore, the 97 basis functions identified as relevant by the RVM approach are not capable of predicting the DFT-PBE ground state set.
This was confirmed by applying the cone-finder algorithm introduced in Section \ref{sec:finding_cone}, which relies on the gradient of the masking function $\eta(\vec{w})$ to find a point inside a target ground state cone in ECI space.
Other basis functions in addition to those of the RVM are therefore needed to endow the cluster expansion with sufficient flexibility such that it can predict the DFT-PBE ground state set of the BCC Li$_x$Al$_{1-x}$ alloy. 

 
To identify a minimal basis set that is both relevant and replicates the DFT-PBE ground states, we started with 505 candidate basis functions that included all 97 RVM basis functions. 
This basis set is sufficiently large to enable replication of the DFT-PBE ground state set as was confirmed with the cone-finder algorithm of Section \ref{sec:finding_cone}.
A posterior distribution was then constructed for the ECI of the 505 cluster basis functions using a prior distribution of the form of Eq. \ref{eqn:prior_general} in which $\gamma \rightarrow \infty$. 
This prior distribution restricts the posterior distribution of the ECI vectors $\vec{w}$ to the DFT-PBE ground state set cone. 
We refer to this posterior distribution as the cone-restricted distribution. 
By construction, any ECI vector $\vec{w}$ sampled from the cone-restricted distribution predicts the correct ground states. 

While ground state replication is ensured with the cone-restricted distribution, a cluster expansion model with 505 basis functions is too large and impractical for downstream Monte Carlo calculations of thermodynamic properties. 
Hence, as a next step, we pruned the 505 cluster basis functions as follows. 
Each of the 408 (= 505-97) non-RVM basis functions was given a removal penalty. 
When the removal of a non-RVM basis function from the original 505 set would prevent replication of the DFT-PBE ground state set, its removal penalty was set to $\infty$. 
When the removal of a non-RVM basis function did not prevent replication of the DFT-PBE ground state set, its penalty was set equal to the change in RMSE as a result of its removal. 
All basis functions with a removal penalty below a particular threshold (i.e. 3.6 meV/atom) were removed. 
For the BCC Li$_x$Al$_{1-x}$ alloy, this resulted in a trimmed set of 214 basis functions consisting of constant and point terms, 8 two-body, 21 three-body, 57 four-body, 55 five-body and 71 six-body clusters.
A new cone-restricted posterior distribution was then constructed for this trimmed basis set. 
The cone-restricted posterior MAP cluster expansion has an RMSE of 3.5 meV per atom between training and predicted formation energies, which is slightly higher than the 3.3 meV/atom RMSE of the RVM MAP cluster expansion. 

In addition to the RVM posterior distribution and the cone-restricted distribution, it is also of interest to analyze the likelihood distribution for the 214 basis functions of the cone-restricted distribution. 
This likelihood distribution is equivalent to a Bayesian posterior distribution, Eq. \ref{eqn:bayes_theorem_contextualized}, having a uniform prior distribution for the ECI of the 214 basis functions (i.e. all prior precisions $\alpha_c=0$). 
The maximum likelihood cluster expansion corresponds to a least squares fit to the training data using the 214 basis functions. 
The RMSE of the maximum likelihood cluster expansion is 2.9 meV/atom, but, similar to the RVM, does not replicate the DFT-PBE ground states.

\begin{figure}[!h]
    \subfloat[]{\includegraphics[width=0.45\textwidth]{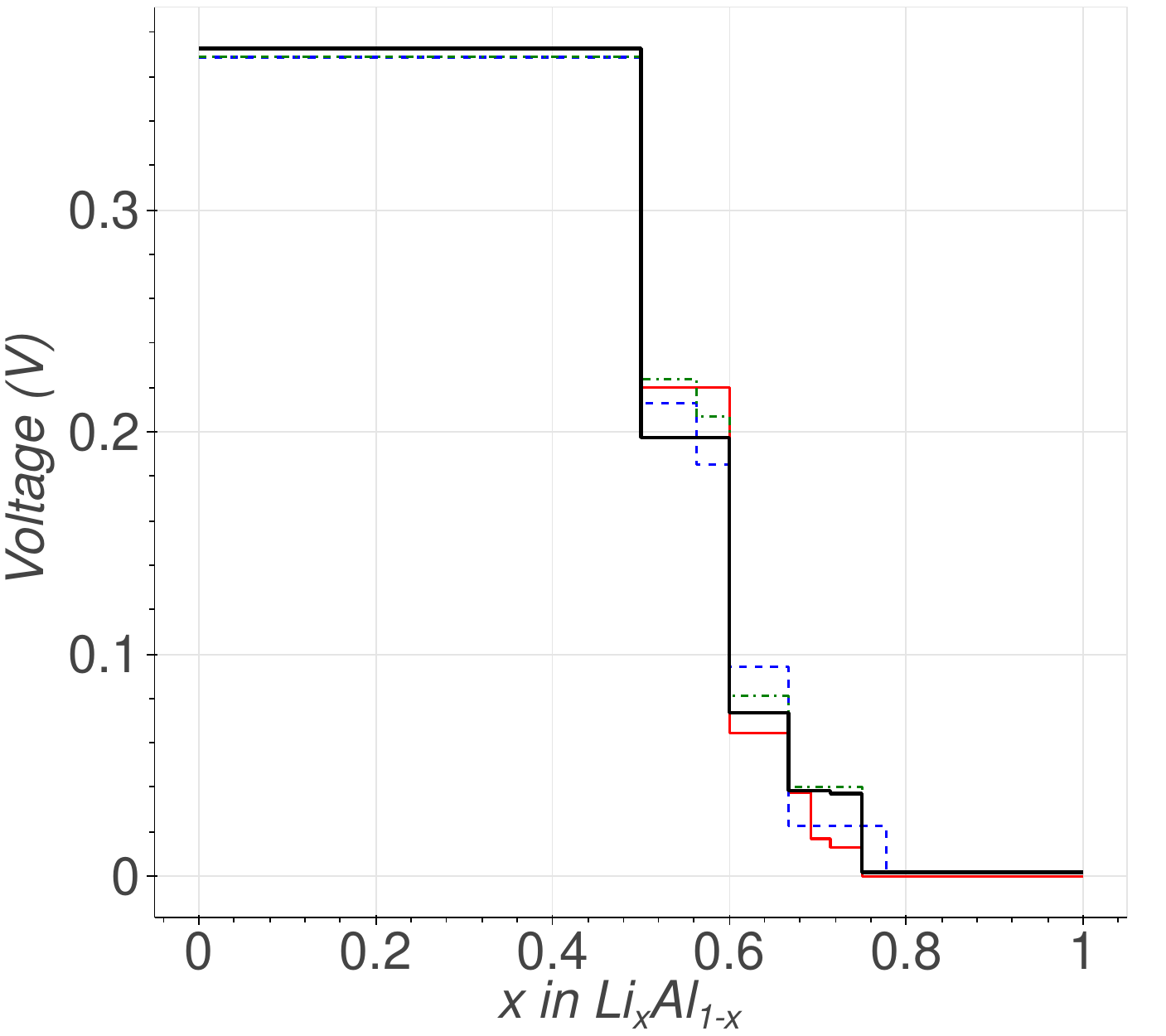}}\\
    \subfloat[]{\includegraphics[width=0.45\textwidth]{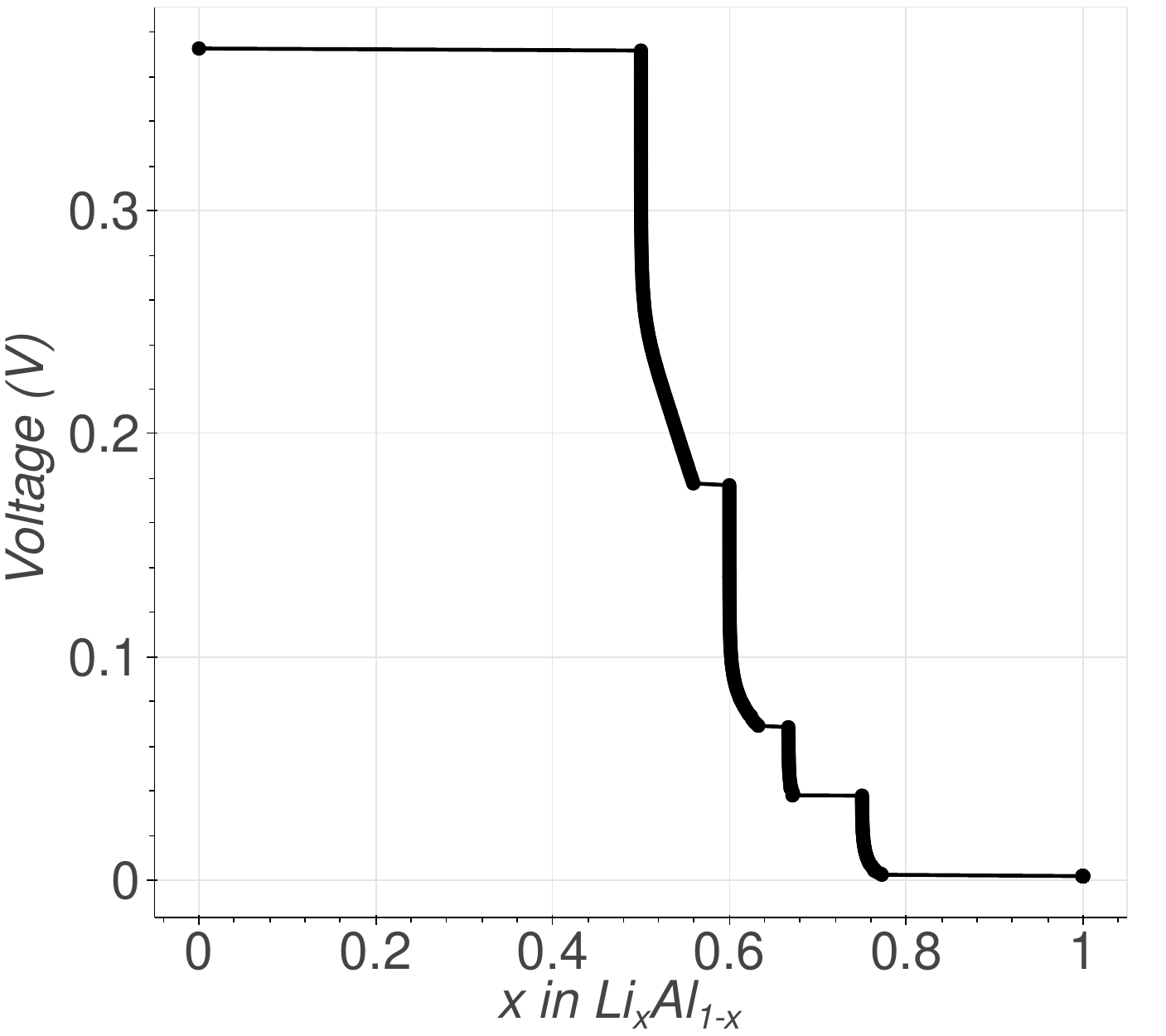}}
    \caption{(a) Zero Kelvin voltage curve for Li$_x$Al$_{1-x}$ predicted by DFT (solid red), the cone-restricted model (solid black), the RVM (dashed blue) and ordinary least squares (dot-dashed green). Only the cone-restricted model replicates all steps of the DFT voltage curve, without any spurious steps. (b) Li$_x$Al$_{1-x}$ voltage curve at 300 K, predicted through thermodynamic Monte Carlo simulation, using a cone-restricted ECI vector. All ground state orderings observed in DFT are preserved, except for the orderings at $x=\frac{9}{13}\sim0.69$ and $x=\frac{5}{7}\sim0.71$ that have already disordered by 300 K.} 
    \label{fig:LiAl_voltage}
\end{figure}

The failure of the RVM MAP and the maximum likelihood cluster expansions to predict the correct ground states manifests itself in downstream properties such as temperature-composition phase diagrams and the electrochemical voltage profile. 
For example, Figure \ref{fig:LiAl_voltage}(a) compares the electrochemical voltage curve at zero Kelvin as calculated with DFT-PBE (red curve) to those predicted with the maximum likelihood (dot-dashed green), the RVM MAP (blue) and the cone-restricted MAP (black) cluster expansions. 
Each step in the zero-Kelvin voltage curve signifies the stability of a ground state ordering.\cite{van2020rechargeable} 
The DFT-PBE voltage curve, and that calculated with the cone-restricted MAP cluster expansion exhibit steps at the same compositions because they both predict the same set of ground states. 
The slight discrepancies in the plateaus, corresponding to the voltages at which one ground state ordering transitions into another ground state ordering, are a reflection of the quantitative accuracy of the cone-restricted MAP cluster expansion. 
The maximum likelihood cluster expansion (i.e. the least squares solution) and the RVM MAP cluster expansion, in contrast, exhibit additional steps and missing steps when compared to the DFT-PBE voltage curve, resulting in not only quantitative discrepancies but also qualitative disagreements. 

An advantage of the cone-restricted posterior distribution is that finite temperature properties predicted with sampled cluster expansions are in qualitative agreement with the zero Kelvin DFT-PBE predictions. 
Figure \ref{fig:LiAl_voltage}(b) shows the voltage curve at 300 K of the BCC Li$_x$Al$_{1-x}$ alloy as calculated by applying Monte Carlo simulations to the cone-restricted MAP cluster expansion. 
The steps corresponding to each ground state ordering show some concentration dependence due to the presence of anti-site defects that are entropically stabilized at finite temperature. 
(Note that, for simplicity, we did not account for vacancies in the Li--Al alloy. However, vacancies are favored on the Li sublattice in the B32 LiAl phase at $x=0.5$. These can be accounted for with a cluster expansion that explicitly treats vacancies in addition to Li and Al.\cite{behara2025crucial})


\begin{figure}[!h]{\includegraphics[width=0.45\textwidth]{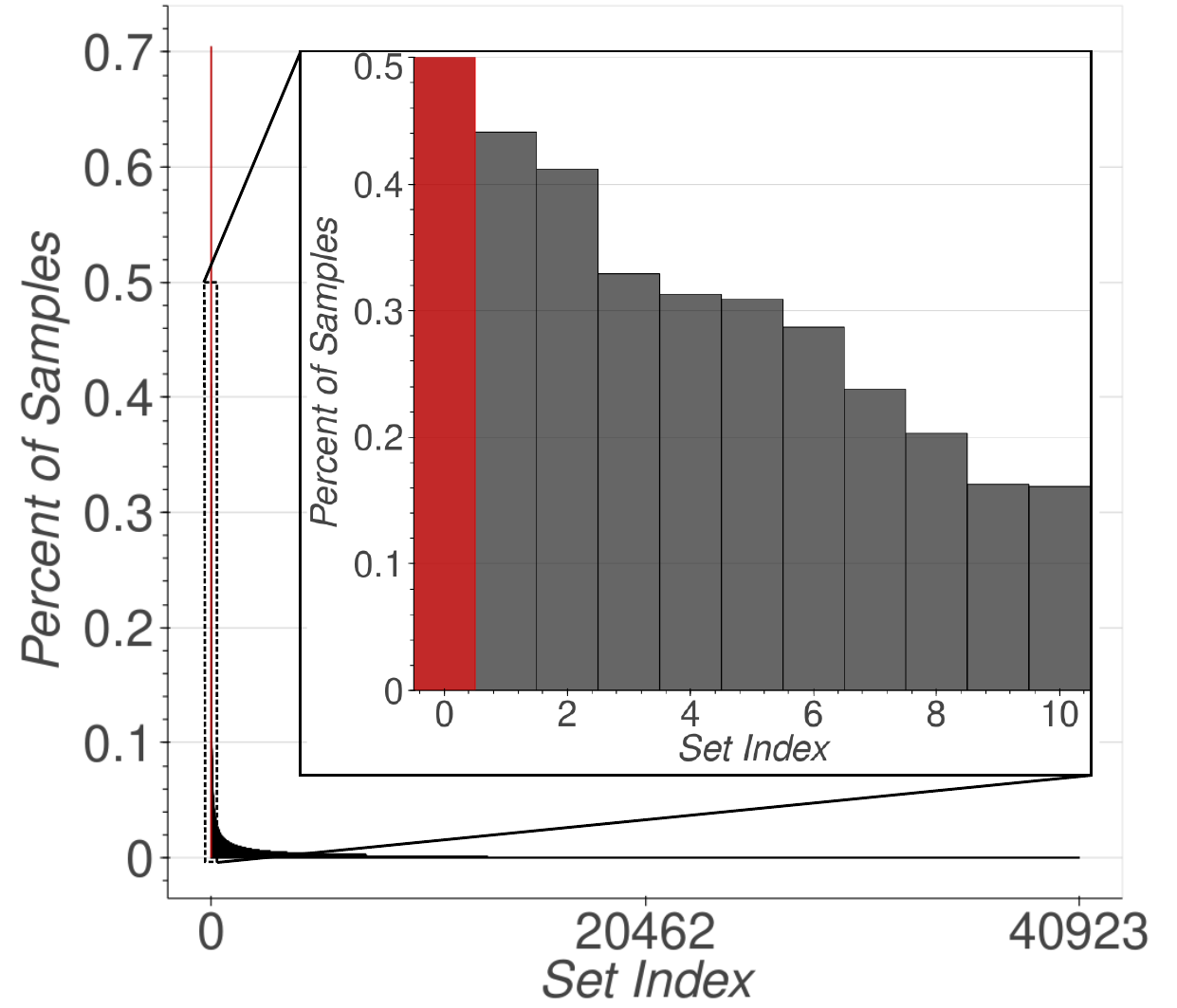}}
    \caption{Unique convex hull sets, and sample frequency, as a percentage of 500,000 samples from the Likelihood function. The histogram bin for the cone containing the ordinary least squares solution is colored in red. In the 500,000 samples, there were 40,923 unique convex hull sets. None matched the DFT-observed ground state set.} 
    \label{fig:cone_frequencies}
\end{figure}

Unlike the cone-restricted posterior distribution, which has finite probabilities only for cluster expansions that predict the same ground state set, cluster expansions sampled from the likelihood do not necessarily all predict the same ground state set.
Figure \ref{fig:cone_frequencies}, for example, shows a ground state set histogram, constructed by sampling 500,000 ECI vectors $\vec{w}$ from the likelihood distribution. 
Each bar in Figure \ref{fig:cone_frequencies} represents the frequency with which a distinct ground state set is predicted by a sampled ECI vector.
Although probability density is highly concentrated across a small subset of the possible ground state sets, the total probability is spread over so many possible sets that even the set corresponding to the maximum likelihood solution (colored in red) still only accounts for approximately 0.7\% of the 500,000 samples. 
Furthermore, there were 40,923 unique ground state sets among the 500,000 sampled cluster expansions $\vec{w}$. 

\begin{figure}
    \centering
    \includegraphics[width=1\linewidth]{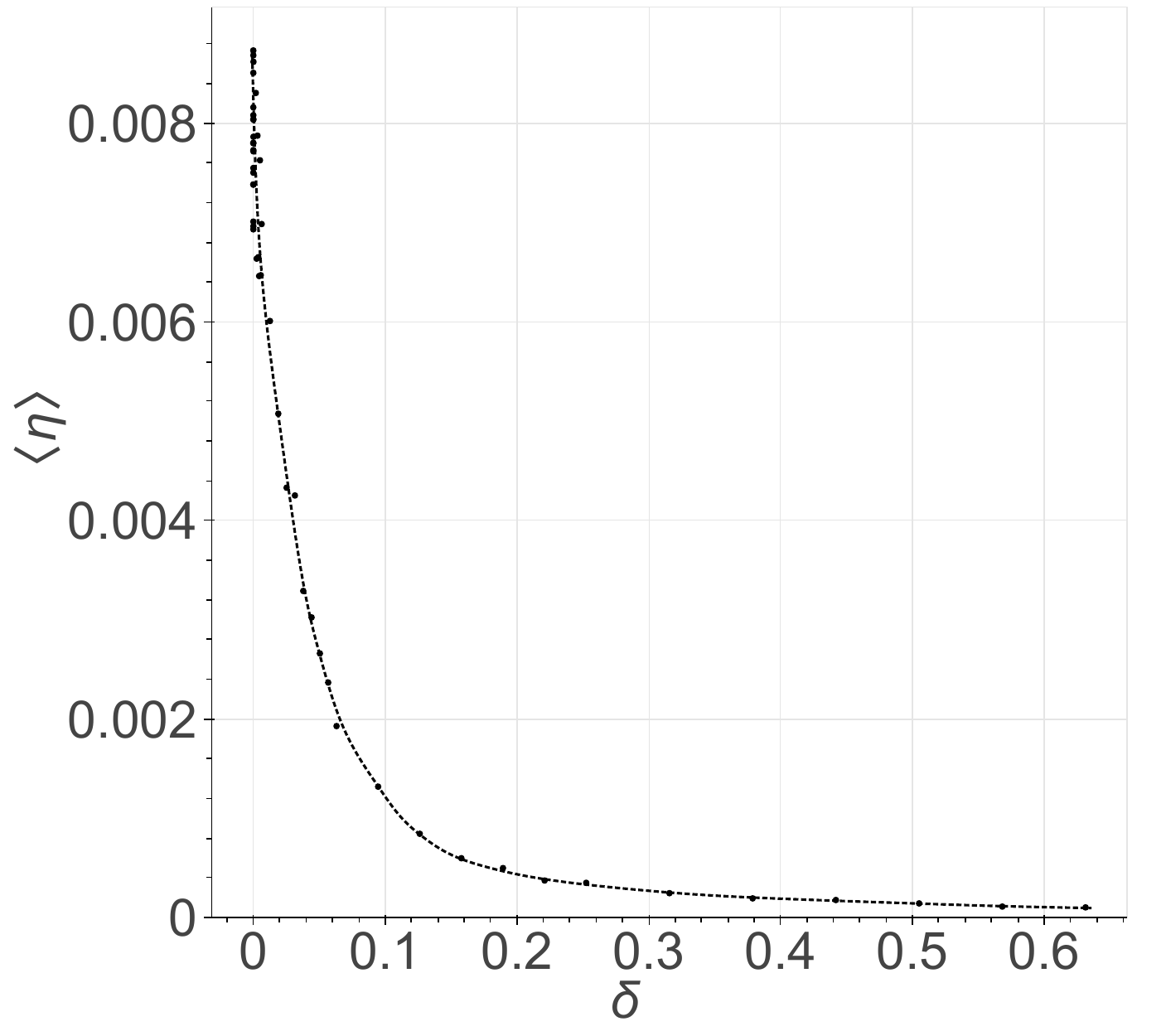}
    \caption{Expectation of the ground state masking function  $\langle \eta (\vec{w})\rangle$ as a function of the conjugate parameter $\delta$. A dashed line is provided as a visual guide.}
    \label{fig:masking_function_vs_conjugate}
\end{figure}

While the DFT-PBE ground state set was never observed among these 500,000 cluster expansions sampled from the likelihood distribution, the space spanned by the 214 basis functions does contain the DFT-PBE ground state cone, meaning that it is still theoretically possible. 
As described in Section \ref{sec:hypothesis_testing}, it is possible to estimate the probability of residing within the DFT-PBE ground state cone relative to the full ECI space according to a particular posterior distribution using Equations \ref{eq:model_probability} and \ref{eq:free_energy_integral}. 
Evaluation of Eq. \ref{eq:free_energy_integral} requires the calculation of the average of the masking function, $\left<\eta(\vec{w})\right>(\delta)$ as the hyperparameter $\delta$ is varied from 0 to $\infty$. 
This average was estimated with Monte Carlo simulations using the likelihood distribution (which corresponds to a posterior distribution in which all $\alpha_c=0$) over a range of $\delta$ (Figure \ref{fig:masking_function_vs_conjugate}). 
The integration of $\left<\eta(\vec{w})\right>(\delta)$ over a finite interval of $\delta$, as opposed to the full interval between $0$ and $\infty$, yields an underestimate of $\Lambda(\delta = \infty)-\Lambda(\delta = 0)$, and thereby overestimates the probability of residing within the ground state cone, Eq. \ref{eq:model_probability}. 
A value of  3.20$\times 10^{-12}\%$ is obtained for the probability of residing in the ground state cone according to the likelihood distribution. 
This result, along with the histogram of Figure \ref{fig:cone_frequencies}, demonstrates that the probability of residing within a particular ground state cone in high-dimensional ECI spaces is very small. 


\section{Discussion}

In this work, we analyzed strategies for constructing Bayesian cluster expansion surrogate models used in first-principles statistical mechanics calculations of materials properties. 
A cluster expansion expresses the energy of interacting atoms as a sum of weights, each multiplied by a cluster basis function. 
Each basis function in turn depends on variables describing a particular set of atomic or electronic degrees of freedom (e.g. chemical site occupancy,\cite{sanchez1984generalized,de1994cluster} atomic displacements,\cite{van2002effect,thomas2013finite} orientations of local magnetic moments,\cite{drautz2004spin,kitchaev2020mapping} molecular orientations \cite{thomas2018hamiltonians} etc.).
We focused on the alloy cluster expansion \cite{sanchez1984generalized} due its importance for modeling configurational disorder in multi-component crystals and its relative simplicity when compared to other cluster expansion surrogate models. 
Many of the conclusions of this study can be generalized to other classes of cluster expansion surrogate models.

Bayesian approaches to parameterizing surrogate models generate probability distributions for the surrogate model weights, $\vec{w}$. 
An attractive feature of a Bayesian approach is the ability to exploit prior knowledge about the weights.\cite{Mueller2009,ober2024thermodynamically} 
This flexibility introduces unknown hyperparameters. 
For the simplest prior distributions, one class of hyperparameters are the precisions $\vec{\alpha}=(\alpha_0,\dots,\alpha_c,\dots)$ of the weights $\vec{w}=(w_0,\dots,w_c,\dots)$. 
Another hyperparameter is the precision of the likelihood function, $\beta$, where $1/\sqrt{\beta}$ reflects the noise on the training data and truncation errors. 
While the hyperparameters themselves have their own prior probability distributions and can thereby be marginalized out, it is common to choose a fixed value for each hyperparameter as guided by one of several principles.\cite{Bishop2006,tipping2001sparse}  
One such guiding principle, often used when constructing cluster expansions,\cite{van2002automating,zarkevich2004reliable,hart2005evolutionary,blum2005using} is to choose hyperparameters that minimize the cross-validation error on a test data set. 
Another approach, referred to as the evidence approximation, chooses values for the hyperparameters that maximize the marginalized likelihood distribution.\cite{tipping2001sparse,ALDEGUNDE2016} 
Within both approaches, further approximations can be made with respect to truncation schemes and whether or not each hyperparameter of the prior distribution is varied independently. 

Historically, cluster expansions have been constructed by minimizing a cross-validation error using a single precision $\alpha$ for all non-truncated cluster basis functions. 
This approach is referred to as Ridge Regression when the prior distribution is normal (i.e. $p=2$ in Eq. \ref{eqn:prior1}), and LASSO when the prior distribution is Laplacian (i.e. $p=1$ in Eq. \ref{eqn:prior1}). 
LASSO is popular as it enables aggressive sparsification of the cluster expansion basis, setting many weights to zero.\cite{nelson2013cluster,nelson2013compressive} 
However, LASSO is less practical for uncertainty quantification as the resulting posterior probability distribution is not easily cast in an analytical form, with the Laplacian prior not being conjugate to the Gaussian likelihood function.

The evidence approximation has been used less commonly in constructing cluster expansions. 
It is most convenient to use a normal distribution as the prior ($p=2$) to ensure analytical expressions for the marginalized likelihood and the posterior distribution on the weights $\vec{w}$. 
The evidence approximation is referred to as Bayesian ridge when a common precision $\alpha$ is used for all weights in the prior and the relevance vector machine (RVM) when the precisions of each weight are varied independently to maximize the marginalized likelihood. 
Aldegunde et al.\cite{ALDEGUNDE2016} were the first to apply the RVM in the context of parameterizing cluster expansions. 

Both the CV and evidence methods that rely on a common precision $\alpha$ for each weight require the specification of a truncation scheme. 
This can be done by explicitly enumerating combinations of basis functions, often guided by physical intuition (e.g. increasing number of sites in a cluster and increasing spatial extent of the cluster),\cite{van2002automating,zarkevich2004reliable} or by a genetic algorithm.\cite{hart2005evolutionary,blum2005using} 
The RVM, in contrast, is able to sparsify the cluster expansion without the need to explore different combinations of non-zero weights. 
As revealed by our study of the Li-Mg alloy, the RVM approach generates physically reasonable cluster expansion models, with most non-zero weights corresponding to small and short-range cluster basis functions. 
Furthermore, it leads to a level of sparsification that is similar to that achieved with LASSO (for the Li$_x$Mg$_{1-x}$ training set, RVM has 95 nonzero ECI while LASSO has 72).
Because the posterior distribution on the weights $\vec{w}$ generated by the RVM is a normal distribution and therefore analytical, it is a convenient distribution for quantifying uncertainty. 
In this respect, we used the RVM to analyze the sensitivity of downstream thermodynamic predictions of the Li$_x$Mg$_{1-x}$ alloy to the approximation of DFT used to generate training data. 
This analysis showed that the variation in predictions using different approximations to DFT is significantly larger than the uncertainties due to the use of surrogate models to interpolate/generalize DFT.
Therefore, for first-principles predictions of thermodynamic properties with uncertainty quantification, it is likely important to perform uncertainty quantification across all relevant approximations to DFT, in addition to uncertainty quantification of the surrogate model.

The evidence approximation enables the determination of a value for $\beta$, the precision appearing in the likelihood function. 
This hyperparameter measures the noise on the training data and implicitly accounts for errors that emerge when a cluster expansion is truncated.
Our study of cluster expansions for the formation energy of the BCC Li$_x$Mg$_{1-x}$ alloy using the evidence approximation shows that $1/\sqrt{\beta}$ decreases as more basis functions are added to the expansion. 
This is an indication that the truncation error continues to decrease as more basis functions are added. 
In the limit of the full basis, $1/\sqrt{\beta}$ should converge to the numerical noise on the DFT training data. 
The asymptotic limit for the Li$_{x}$Mg$_{1-x}$ alloy suggests a value of less than 1 meV/atom, which is of the order of the numerical errors of the DFT energies due to k-point sampling. 
The RVM method, which generates a relatively compact set of non-zero weights, has the lowest value of $1/\sqrt{\beta}$.

A major challenge in constructing cluster expansion surrogate models is to ensure that they predict the same ground states as predicted with the higher accuracy first-principles electronic structure method. 
For an alloy cluster expansion, the set of weight vectors, $\vec{w}$, that all predict the same ground state set forms cones emanating from the origin in ECI space. 
Our analysis of the Li$_x$Al$_{1-x}$ alloy showed that sampled cluster expansions from typical posterior distributions (e.g. likelihood, RVM) have a very high degree of variability in the predicted ground state sets. 
This poses difficulties for uncertainty quantification since the sampled cluster expansions do not predict the same qualitative behavior in downstream thermodynamic predictions. 
Within a Bayesian framework, certainty about ground states can be embedded within the prior distribution. 
As shown here, prior distributions expressed in terms of masking functions in ECI space are able to restrict the posterior distribution to a target ground state cone. 
This then ensures that all sampled cluster expansions from the posterior distribution predict the same qualitative behavior and can be used in a meaningful way to calculate uncertainties. 
Furthermore, the probability of a particular ground state set can be calculated using free energy integration techniques of statistical mechanics as described in Section \ref{sec:hypothesis_testing}. 
Monte Carlo algorithms to sample the ground-state-restricted posterior distributions require an initial point within the desired ground-state cone in ECI space.
To identify such an ECI vector, $\vec{w}$, we developed a cone-search algorithm (Section \ref{sec:finding_cone}). 
The cone-search algorithm not only finds points within a target ground state cone, but also indicates whether a particular level of truncation of the cluster expansion is capable of replicating a particular set of ground states.

While the focus here has been on relatively simple alloy cluster expansions, the same methods and analysis can be performed for any surrogate model expressed as a linear expansion in terms of basis functions.
In particular, a similar approach can be implemented to enforce qualitative behavior as exhibited by atomic cluster expansion surrogate models.\cite{drautz2019atomic} 
For example, it may be essential that the atomic cluster expansion accurately predicts the relative stability between different crystal structures.
Similar masking functions and prior distributions as introduced here for the ground state problem of the alloy cluster expansion can be used to generate posterior distributions on atomic cluster expansions that guarantee a particular set of qualitative predictions.

\section{Conclusion}
In this work, we analyzed strategies for constructing Bayesian cluster expansions. 
Bayesian hyperparameter selection was addressed using cross-validation and the evidence approximation, using the BCC Li$_x$Mg$_{1-x}$ alloy as a model. 
The Relevance Vector Machine (RVM) variant of the evidence approximation is found to be a highly effective method, producing posterior distributions that quantify cluster expansion uncertainty, with low root-mean-squared error (RMSE). 
The RVM automatically performs truncation and outperforms explicit truncation schemes with respect to model evidence and model error. 
The uncertainty across DFT approximations was also analyzed and compared to the uncertainty of the cluster expansion model.
Confidence regions on predicted results for each DFT approximation show that the uncertainty between DFT approximations is larger than the uncertainties associated with cluster expansions. 
Finally, the problem of ground state replication was addressed in a study of the Li$_x$Al$_{1-x}$ system. 
Cluster expansion models that predict a given set of ground state configurations all reside in a continuous ray-bounded domain of the model parameter space. 
As both criteria are difficult to satisfy with sampling, 
we developed a ground state cone-search algorithm to determine cluster expansion models that replicate a target ground state set. 

\section{Acknowledgments}


This work was supported by the U.S. Department of Energy, Office of Basic Energy Sciences, Division of Materials Sciences and Engineering under Award \#DE-SC0008637 as part of the Center for Predictive Integrated Structural Materials Science (PRISMS Center).
The research reported here made use of computational resources of the Center for Scientific Computing (CSC) at UCSB and of the National Energy Research Scientific Computing Center (NERSC), a Department of Energy Office of Science User Facility using NERSC award BES-ERCAP0026626.
Research was also carried out at the Center for Functional Nanomaterials, Brookhaven National Laboratory, through the U.S. Department of Energy, Office of Basic Energy Sciences, contract DE-AC02-98CH10866, under award CFN312109.
The CSC is supported by the California NanoSystems Institute and the Materials Research Science and Engineering Center (MRSEC; NSF DMR 2308708) at UC Santa Barbara.
Computational facilities administered by CSC were purchased with funds from the National Science Foundation (CNS-1725797).

\clearpage
\appendix

\section{Gradient of the masking function in high-dimensional composition spaces}
\label{app:gradient_masking_function}

The gradient of the masking function, $\eta(\vec{w})$, with respect to the ECI vector $\vec{w}$, for a binary system (Eq. \ref{binary_trapezoid}) can be generalized to high-dimensional composition spaces.  
The integral of the piecewise linear function connecting the energies of the target ground states in high-dimensional composition spaces can be written as
 \begin{equation}
     \int f(\vec{w},\vec{x})dx = \frac{1}{d+1}\sum_{s\in S} \left( \sum_{i \in s} \vec{\xi_i}\vec{w} \right) \frac{1}{d!}|det(\text{X}_s)|
 \end{equation}
where $d$ is the dimension of the composition space and $s$ indexes simplexes in composition space whose vertices are the compositions of the target ground states, $i$. 
In the above integral, $\text{X}_s$ is a square matrix whose columns are made up of composition vectors of each vertex with an additional 1 appended after the last element in each column vector.
A similar expression can be formulated for the integral of the piece-wise linear function $g(\vec{w},\vec{x})$, corresponding to the convex hull for the ECI vector $\vec{w}$.
%
The negative gradient direction in ECI space then becomes
\begin{equation}\label{eq:neg_grad}
     -\nabla_w \eta(\vec{w}) \propto  \sum_{s\in S}|\det(\text{X}_s)| \sum_{i \in s} \vec{\xi_{i}}  - \sum_{s'\in S'}|\det(\text{X}_{s'})| \sum_{i' \in s'} \vec{\xi_{i'}} 
\end{equation}
where the primed (unprimed) quantities come from the integral of $f$ ($g$).  

The definition of $\eta(\vec{w})$ in Eq. \ref{eqn:continuous_gsa_first} enforces a strict match between the predicted and the target ground states, being zero only if all the ground states as predicted by the ECI vector $\vec{w}$ are identical to the desired ground states. 
Other definitions for $\eta(\vec{w})$ that are not as stringent can also be formulated. 
For example, there may be certainty about a subset of the DFT predicted ground states, but not others. 
A ``softened" definition can take the form
\begin{equation}\label{eq:alternative_gs_metric}
 \sum_{i \in h} [\vec{\xi}^T_i \vec{w} - g(\vec{w},x)] + \sum_{j \in h'}[f(\vec{w},x) - \vec{\xi}^T_j\vec{w}]    
\end{equation}
where $h$ is the set of all imposed ground states, and $h'$ is the set of all points with formation energies less than $f(\vec{w},x)$. 
The function can be constructed only to consider specific composition intervals, or certain subsets of $h$ and $h'$. 
These ``softened" definitions enforce unions of ground-state sets. 
\clearpage
\bibliography{citations}
\end{document}